\documentclass[12pt]{article}

\pdfoutput=1
\usepackage{graphicx}
\usepackage{amssymb}
\usepackage{amsmath}
\usepackage{color}
\usepackage{cite}
\usepackage{subfigure}
\usepackage[colorlinks=true,linkcolor=blue,citecolor=blue]{hyperref}

\begin{document}

\newcommand{\be}{\begin{equation}}
\newcommand{\ee}{\end{equation}}


\begin{titlepage}

\begin{flushright}
ICRR-Report-649-2012-38  \\
IPMU 13-0100
\end{flushright}

\vskip 2cm

\begin{center}

{\Large \bf
	Gravitational waves from a curvaton model \\[2mm]
	 with blue spectrum
}

\vspace{1cm}

{Masahiro Kawasaki$^{(a, b)}$, Naoya Kitajima$^{(a)}$ and Shuichiro Yokoyama$^{(a)}$}

\vskip 1.0cm

{\it
$^a$Institute for Cosmic Ray Research,
     University of Tokyo, Kashiwa, Chiba 277-8582, Japan\\
$^b$Kavli Institute for the Physics and Mathematics of the Universe (WPI), 
Todai institute for Advanced Study, University of Tokyo, Kashiwa, Chiba 277-8568, Japan\\
}

\vskip 1.0cm

\begin{abstract}
	We investigate the gravitational wave background induced by the first order scalar perturbations in the curvaton models. 
	We consider the quadratic and axion-like curvaton potential which can generate 
	 the blue-tilted power spectrum of curvature perturbations on small scales and
	derive the maximal amount of gravitational wave background today.
	We find the power spectrum of the induced gravitational wave background has a characteristic 
	peak at the frequency corresponding to the scale reentering the horizon at the curvaton decay, 
	in the case where the curvaton does not dominate the energy density of the Universe.
	We also find the enhancement of the amount of the gravitational waves in the case
	where the curvaton dominates the energy density of the Universe.
	Such induced gravitational waves would be detectable 
	by the future space-based gravitational wave detectors or pulsar timing observations.
\end{abstract}

\end{center}

\end{titlepage}

\newpage

\vspace{1cm}

\section{Introduction} \label{intro}

Recent cosmological observations such as Planck \cite{Ade:2013lta} or Wilkinson 
Microwave Anisotropy probe (WMAP) \cite{Hinshaw:2012fq} strongly supports the existence of 
an accelerated expansion era called inflation in the very early stage of the universe.
In addition to solving some difficulties in the standard Big Bang cosmology 
such as the horizon problem, inflation can naturally generate the primordial 
seeds of density perturbations on superhorizon scale by expanding the quantum 
fluctuations of some scalar field on very small scales.
This mechanism for generating the density perturbations generally predicts 
the almost scale-invariant spectrum of curvature perturbations obeying the Gaussian 
statistics,which is consistent with the observed temperature anisotropies of the cosmic 
microwave background (CMB).
Similarly, the standard inflationary paradigm  predicts the generation of the primordial tensor metric 
perturbations which may be detected as the gravitational wave background (GWB) by future observations.

The inflation is driven by some scalar field called inflaton whose potential is nearly flat.
Although various inflation models have been proposed so far, unfortunately, 
there is no promising candidate for an inflaton.
Moreover, the density perturbation may be produced by another scalar field like the curvaton model \cite{Enqvist:2001zp,Lyth:2001nq,Moroi:2002rd}.
One of the difficulty to identify the model is due to lack of clues 
from the observations on small scales.
While the CMB observation is a powerful tool for constraining the spectrum of 
the curvature perturbations on large scales, we  know little about the one 
on small scales, so we have only a little information to constrain the inflation model.
On the other hand, future and current detectors of gravitational waves as shown later
have a high sensitivity at higher frequency modes corresponding to the smaller scale fluctuations
than that observed by CMB anisotropies.
Thus, it is interesting to investigate the signal of the primordial gravitational waves on small scales, 
which would be a powerful tool to reveal the inflationary dynamics over a long period
 in combination with the CMB observations.

There are several  mechanisms for generating the primordial GWBs.
First, as mentioned before, they are generated by the quantum fluctuations
of the tensor metric perturbations in the inflationary era, which typically has an almost
scale-invariant power spectrum, and the amplitude of such gravitational waves
is typically given by the inflationary Hubble scale.
In addition to the direct detections, 
it is also expected to detect the signal of gravitational waves on cosmological scales
through the observations of the B-mode polarization of the CMB anisotropy
\cite{Seljak:1996gy,Kamionkowski:1996zd,Smith:2005mm}. 
Second, the gravitational waves can be also induced by primordial scalar perturbations.
In principle, the scalar and tensor perturbations evolve independently and are not mixed 
at linear order on the homogeneous
isotropic FLRW universe, but at the second order
these perturbations are not independent any more
~\cite{Matarrese:1992rp,Matarrese:1997ay,Noh:2004bc,Carbone:2004iv,Nakamura:2004rm}.
Hence the stochastic GWB can be sourced from the quadratic component of the first order scalar perturbations
in the metric and energy-momentum tensor~\cite{Mollerach:2003nq,Ananda:2006af,Baumann:2007zm}.
Actually,  the blue-tilted adiabatic curvature perturbations generate a large amount 
of gravitational waves at high frequencies, 
which would be detectable by future observations~\cite{Assadullahi:2009jc,Alabidi:2012ex,Alabidi:2013lya}
Since such curvature perturbations have the potential to form a large number of  primordial black holes (PBHs),
the amount of the scalar-induced gravitational waves 
is also constrained by the PBH formation~\cite{Saito:2008jc,Bugaev:2009zh}.

In this paper, we investigate the gravitational wave generation in the curvaton model.
Various relevant studies have been done in the literature~\cite{Bartolo:2007vp,Enqvist:2008be,Assadullahi:2009nf,Suyama:2011pu}.
We consider two cases with the quadratic and also the axion-like potential,
both where the blue-tilted adiabatic curvature perturbations can be easily realized.
As is well-known, such blue-tilted spectrum is not compatible with the current CMB observations 
such as Planck  \cite{Ade:2013lta}.
Hence, here we also introduce the inflaton fluctuations as a source of
the adiabatic curvature perturbations on large scales  so that we realize the COBE normalization and
the slightly red-tilted power spectrum of the curvature perturbations.
On smaller scales, the blue-tilted component sourced from the curvaton fluctuations
is dominated. 
In this set-up, we investigate the GWB induced from not only the scalar adiabatic perturbations 
but also from the transverse-traceless part of the energy momentum tensor
which is due to the kinetic term of the curvaton field ~\cite{Bartolo:2007vp}. 
We also study both cases where the energy density of the curvaton field is dominant 
or still subdominant in the Universe at the curvaton decay~\cite{Assadullahi:2009nf}\footnote{
	Here, we consider the case with a single curvaton field. 
	In~\cite{Suyama:2011pu}, the authors have considered  two-curvaton scenarios and have investigated 
	the generation of the scalar-induced gravitational waves during the era when one of the curvaton is dominant.
}.
Then, we discuss the detectability of such scalar-induced GWB in the future experiments:
such as the space-based gravitational wave detectors e.g., Laser Interferometer 
Space Antenna (LISA)~\cite{LISA_NASA}, DECi-hertz Interferometer Gravitational wave 
Observer (DECIGO)~\cite{Kawamura:2006up,DECIGO} and Big Bang Observer 
(BBO)~\cite{Corbin:2005ny}, and also
the pulsar timing observations \cite{Lommen:2002je} like Square Kilometre Array (SKA) \cite{SKA}.

The remainder of this paper is organized as follows.
In section \ref{curvaton_model}, we briefly review the curvaton model and show the
blue-tilted power spectrum in the quadratic and axion-like model.
In section \ref{induced_GW}, we formulate the GWB induced from the scalar adiabatic perturbations
and also the transverse-traceless component of the energy momentum tensor due to
the kinetic term of the curvaton. In section \ref{sec;results}, 
we show the resultant power spectrum of the GWB in addition to the sensitivity curves of the future
gravitational wave detectors. 
Then, we conclude in section \ref{conc}.


\section{Blue-tilted curvature perturbations in curvaton model} 
\label{curvaton_model}

In this section, we briefly review curvaton scenarios which could produce
the blue-tilted power spectrum of the curvature perturbations.
In order not to conflict with the current cosmological observations which
disfavor the blue-tilted power spectrum,
we consider a mixed scenario where the primordial curvature perturbations
have been sourced from not only the curvaton but also the inflaton fluctuations
with slightly red-tilted power spectrum.

\subsection{Basics of the curvaton scenario} 
\label{basics}

In the curvaton scenario \cite{Enqvist:2001zp,Lyth:2001nq,Moroi:2002rd},
one introduces a scalar field besides inflaton, so-called curvaton denoted 
as $\sigma$, which is a subdominant component and acquires quantum fluctuations 
during inflation.
In general, the curvaton slowly rolls down its own potential during inflation 
because its mass is smaller than the Hubble parameter and it starts to oscillate 
coherently when the Hubble parameter $H$ becomes comparable to the mass after inflation.
Assuming the quadratic potential, such coherently oscillating curvaton  behaves like a pressureless matter 
whose energy density decreases obeying $\propto a^{-3}$ where $a$ is the scale factor 
of the cosmic expansion.
Hence, in the radiation dominated Universe where the energy density of the Universe 
evolves as $\propto a^{-4}$, the ratio of the energy density of the curvaton to 
that of the radiation  relatively increases and the isocurvature perturbations 
sourced from the curvaton quantum fluctuations contributes to the evolution of 
the adiabatic curvature perturbations on super-horizon scales. 
After the curvaton decays into the radiation when the Hubble parameter becomes 
equal to the decay rate of the curvaton written as $\Gamma_\sigma$,
the resultant adiabatic curvature perturbations stay constant in time.

Based on the $\delta N$ formalism \cite{Starobinsky:1986fxa},
the curvature perturbations on the uniform energy density hypersurface
on super-horizon scales are given by the perturbation of the e-folding number
and can be expanded in terms of the fluctuations of the fields as
\be
	\zeta = \delta N = N_{\phi} \delta\phi_* + N_\sigma \delta\sigma_* + \cdots,
\ee
where $\phi$ and $\sigma$ denote the inflaton and curvaton respectively, 
$N_\phi = \partial N / \partial \phi_*$ and $N_\sigma = \partial N / \partial \sigma_*$
with $N = \int_{t_*}^t H dt$.
Here,  an subscript $*$ denotes a value at the initial flat hypersurface
and we take $t=t_*$ to be a time when the scale of interest exits the Hubble horizon 
during inflation.
Assuming the standard single slow-roll inflation and
the quadratic potential for the curvaton during the oscillating phase,
we have \cite{Kawasaki:2011pd}
\be
	N_\phi = \frac{1}{\sqrt{2\epsilon M_P^2}} ~~~\text{and}~~~ 
	N_\sigma = \frac{2r(\eta)}{4+3r(\eta)}
	\frac{d\ln \sigma_{\rm osc}}{d\sigma_*},
\ee
where $M_P = (8\pi G)^{-1/2}$ with $G$ being Newton's constant is the reduced 
Planck mass, $\epsilon = - \dot{H}/H^2$ is a slow-roll parameter during inflation (overdot denotes the time derivative), 
and a subscript $ ``{\rm osc}"$ denotes a value at the start of the curvaton 
oscillation.
Here, we also consider the curvature perturbations before the curvaton decays
and hence
$r(\eta)$ is the time-dependent ratio of the energy density of the curvaton 
$\rho_\sigma$ to that of radiation $\rho_r$ given by
\be
	r(\eta) = 
	\begin{cases}
		\frac{\rho_\sigma(\eta)}{\rho_r(\eta)} ~~~
		&\text{(before the curvaton decay)} \\[2mm]
		r_D \equiv \frac{\rho_\sigma}{\rho_r} \big|_{t=t_D} ~~~
		&\text{(after the curvaton decay)},
	\end{cases}
\ee
where a subscript ``$D$" denotes a value when the curvaton decays into radiation.
Because of $r(\eta) \propto a $ before the curvaton decay, 
in the case where the \textit{first} reheating induced by the inflaton decay
occurs before the beginning of the curvaton oscillation, 
$r_D$ is calculated as
\be
	r_D = \frac{1}{6} \bigg( \frac{\sigma_{\rm osc}}{M_P} \bigg)^2 \bigg( \frac{m_\sigma}{\Gamma_\sigma} \bigg)^{1/2} ~~~\text{for}~~~
	m_\sigma  < \Gamma_\phi,
	\label{r_D_1}
\ee
where 
$\Gamma_\phi$ and $\Gamma_\sigma$ are respectively 
the decay rate of the inflaton and curvaton.
In another case where 
$m_\sigma > \Gamma_\phi$,
that is,
the curvaton starts to oscillate before
the inflaton decays into radiation,
we obtain
\be
	r_D = \frac{1}{6} \bigg( \frac{\sigma_{\rm osc}}{M_P} \bigg)^2 \bigg( \frac{\Gamma_\phi}{\Gamma_\sigma} \bigg)^{1/2} ~~~\text{for}~~~
	m_\sigma > \Gamma_\phi.
	\label{r_D_2}
\ee

Then, the power spectrum of the curvature perturbations is obtained by 
\be
    \langle \zeta({\bf k},\eta) \zeta({\bf k}',\eta)\rangle 
    =  \delta^{(3)}({\bf k} + {\bf k}') 
    {2 \pi^2 \over k^3} \mathcal{P}_\zeta (k,\eta),
\ee
\be
	\mathcal{P}_\zeta (k,\eta) = \mathcal{P}_{\zeta,{\rm inf}}(k) 
	+ \mathcal{P}_{\zeta,{\rm curv}}(k),
	\label{P_zeta_total}
\ee
where $\mathcal{P}_{\zeta,{\rm inf}} (k,\eta)$ and $\mathcal{P}_{\zeta,{\rm curv}}(k)$
are contributions from the inflaton and curvaton,
respectively, and they are given by
\begin{eqnarray} 
    \mathcal{P}_{\zeta,{\rm inf}}(k) 
    & = &{1 \over 2 \epsilon} \left( {H_{\rm inf} \over 2 \pi M_P}\right)^2,
    \label{power_inf} \\[0.5em]
    \mathcal{P}_{\zeta,{\rm curv}}(k) 
    & = &\bigg( \frac{r(\eta)}{4+3r(\eta)} \bigg)^2
    \left({2 d \ln \sigma_{\rm osc} \over d \sigma_* } \right)^2
    \left( {H_{\rm inf} \over 2 \pi}\right)^2 
    \nonumber \\[0.5em]
    & & \equiv  \bigg( \frac{r(\eta)}{4+3r(\eta)} \bigg)^2 \mathcal{P}_{S,{\rm curv}}(k) ,
    \label{power_curv}
\end{eqnarray}
where $ \mathcal{P}_{S,{\rm curv}}$ is the power spectrum of the isocurvature 
perturbations induced by the curvaton and we have used
\be
   \langle \delta \phi({\bf k})_* \delta \phi ({\bf k}')_* \rangle 
   = \langle \delta \sigma({\bf k})_* \delta \sigma ({\bf k}')_* \rangle 
   =  \delta ({\bf k} + {\bf k}') \left({2 \pi^2 \over k^3}\right)
   \left( {H_{\rm inf}\over 2 \pi} \right)^2.
\ee
Recent cosmological observations have indicated that
the power spectrum of the primordial 
curvature perturbations 
is almost scale-invariant and ${\mathcal P}_\zeta \simeq 10^{-10}$
over the observable scales about
$k \lesssim 1~{\rm Mpc}^{-1}$.
On the other hand, due to the difficulty of the direct observation of
the primordial fluctuations,
${\mathcal P}_\zeta$ on smaller scales
is comparatively free from tight observational bounds,
except for the constraints from
the abundance of PBHs \cite{Carr:2009jm} 
or ultra-compact minihalos (UCMHs) \cite{Josan:2010vn,Bringmann:2011ut},
CMB $\mu$-distortion \cite{Chluba:2012we}
and GWBs \cite{Saito:2008jc}.
Hence, we rely on the contribution from the inflaton fluctuations
for the curvature perturbations on large scales,
which have an almost scale-invariance power spectrum,
and assume that the
curvature perturbations on the smaller scales
 is dominated by the part sourced from
the curvaton fluctuations 
with extremely blue-tilted power spectrum.
We parametrize
the small scale power spectrum of curvature perturbations 
induced from the curvaton fluctuations as
\be
	\mathcal{P}_{\zeta,{\rm curv}}(k) 
	= \mathcal{P}_{\zeta,{\rm curv}}(k_c) \bigg( \frac{k}{k_c} \bigg)^{n_\sigma-1}
	\label{P_zeta_curv}
\ee
where $n_\sigma$ is the spectral index, $k_c$ is some reference value of wave number taken to be $k_c = 1~{\rm Mpc}^{-1}$ and
\be
	\mathcal{P}_{\zeta,{\rm curv}}(k_c)
	= \mathcal{P}_{\zeta,{\rm inf}} = 2 \times 10^{-9}.
\ee
Before the discussion about the spectral index $n_\sigma$
for some models,
we would like to note the non-Gaussianity of the primordial curvature perturbations, 
characterized by the nonlinearity parameter $f_{\rm NL}$, in the case we consider here.
The curvaton model has been well-known as a model which
can predict the large non-Gaussianity such as $f_{\rm NL} \sim 1/r_D \gtrsim 10$.
Very recently, Planck observation placed a stringent constraint on $f_{\rm NL}$, 
giving $-8.9 < f_{\rm NL} < 14.3$ at two sigma level~\cite{Ade:2013lta}. 
However, even if we take small $r_D$ in the later discussion,
our model does not conflict with the Planck constraint. 
Because we rely on the inflaton to generate the large scale curvature perturbations
which are observed by Planck  and $f_{\rm NL}$ is suppressed by roughly 
$(\mathcal{P}_{\zeta,{\rm curv}} / \mathcal{P}_{\zeta,{\rm inf}})^2 \times 1/r_D$.

\subsection{Quadratic curvaton model} \label{quadratic_curvaton}

First, let us consider the curvaton model with quadratic potential,
\be
	V(\sigma) = \frac{1}{2} m_\sigma^2 \sigma^2.
\ee
For this potential, under the slow-roll approximation
the field value of $\sigma$ when the mode $k$ exits the horizon is obtained as
\be
	\sigma_\ast(k) = \sigma_{e} \exp \left[ {m_\sigma^2 \over 3 H_{\rm inf}^2} N_{e}  \right]
	 \bigg( \frac{k}{k_p} \bigg)^{-{m_\sigma^2 \over 3H_{\rm inf}^2}},
\ee
where $\sigma_e$ is the field value at the end of the inflation, $N_e$ is the e-folding number. 
We set $N_e = 50$ and $k_p = 0.002~{\rm Mpc}^{-1}$ 
from the time when the pivot scale $k_p$ leave the horizon to the end of the inflation.
Because of $m_\sigma \sim H_{\rm inf}$ in the present setup, the curvaton starts to oscillate soon after the inflation, which leads to $\sigma_e= \sigma_{\rm osc}$. 
From the above expression,
we have
\be
      { d \ln \sigma_{\rm osc} \over d \sigma_* } = \frac{1}{\sigma_\ast(k)} \propto \bigg( \frac{k}{k_p} \bigg)^{{m_\sigma^2 \over 3H_{\rm inf}^2}},
\ee
and hence the spectral index of the power spectrum of the curvature perturbations
induced from the curvaton fluctuations in this model is given by \cite{Lyth:2001nq}
\be
	n_\sigma \simeq 1 + \frac{2 m_\sigma^2}{3H_{\rm inf}^2},
\ee
where we have neglected the contribution from the time-derivative of Hubble parameter, 
which is $\mathcal{O}(\epsilon)$ and much smaller than $2m_\sigma^2/3H_{\rm inf}^2$, 
because we set $m_\sigma \sim H_{\inf}$ to realize the extremely blue-tilted power spectrum.

\subsection{Axion-like curvaton model} \label{axionic_curvaton}

Here  we briefly introduce the axion-like curvaton model having extremely 
blue-tilted power spectrum.
This model is based on a supersymmetric axion model~\cite{Kasuya:2009up}, 
and it was also discussed in the context of the PBH formation in~\cite{Kawasaki:2012wr}.
Omitting some detailed discussion, the relevant ingredient is a complex scalar 
field $\Phi$ denoted as $\Phi = \varphi e^{i\theta}/\sqrt{2}$ 
which has a charge for some global $U(1)$ symmetry.
Taking into account the supergravity effect, 
the potential for $\varphi$ 
during inflation is given by
\be
	V(\varphi) = \frac{1}{2} c H_{\rm inf}^2 (\varphi - f)^2, \label{eq:bluepot}
\ee
where $f$ is the minimum of $\varphi$ and $c$ is some numerical constant assumed 
to be $\mathcal{O}(1)$.
Note that, since the mass of $\varphi$ is comparable to the Hubble parameter, 
$\varphi$ moves toward the minimum somewhat rapidly  during inflation.
Here and hereafter, we assume that $\varphi$ has a large initial value 
which is much larger than $f$, 
and the phase direction acquires
fluctuations during inflation.
After $\varphi$ reaches the minimum,
the curvaton is defined as the phase component of the complex scalar field: 
$\sigma = f\theta$.
Like an axion model, we assume that the global $U(1)$ symmetry is broken by some nonperturbative effect, which leads the potential of the curvaton given by
\be
	V(\sigma) = \Lambda^4 \bigg[ 1-\cos \bigg( \frac{\sigma}{f} \bigg) \bigg] 
	\simeq \frac{1}{2} m_\sigma^2 \sigma^2
\ee
where $\Lambda$ is some energy scale and the curvaton mass is given by 
$m_\sigma = \Lambda^2/f$ near the minimum.
The interaction of the curvaton with ordinary matter is suppressed by $f$ 
and the decay rate of the curvaton is given by
\be
	\Gamma_\sigma = \frac{\kappa^2}{16\pi} \frac{m_\sigma^3}{f^2},
	\label{Gamma_sigma}
\ee
where $\kappa$ is a dimensionless coupling constant.

Let us consider the generation of the fluctuations of the curvaton in this model.
During inflation, the fluctuation of the phase direction $\theta$ is roughly given by  
$\delta\theta_* \sim H_{\rm inf} / \varphi_*$ at the horizon exit
and it stays constant on super Hubble scales.
After $\varphi$ reaches the minimum,
the field value of the curvaton is unchanged 
because the mass of the curvaton is much smaller than $H_{\rm inf}$
and
 the fluctuation of $\theta$ can be identified with that of the curvaton
as \footnote{
The fluctuations of $\varphi$ also give a 
contribution to $\delta\theta$. 
However, it is small (suppressed by $\theta$) and hence can be neglected.}
\be 
    {d \ln \sigma_{\rm osc} \over d \sigma_*} \delta \sigma_* = 
    {\delta \sigma \over \sigma}\Biggr|_{\varphi = f}
    = {\delta \theta \over \theta_i}\Biggr|_{\varphi=f} = {\delta \theta_* \over \theta_i}
    = {1 \over \varphi_*(k) \theta_i} \left( {H_{\rm inf} \over 2 \pi} \right),
\ee
where $\theta_i$ is the initial misalignment angle 
and we assume $\theta_i > H_{\rm inf}/(2\pi f)$ 
in order for the quantum fluctuations not to dominate over the classical value.
As can be seen in the above expression, the scale dependence of the curvature perturbations 
sourced from the fluctuations of the curvaton
is determined 
by solving the equation of motion for $\varphi$ during inflation.
For the potential given by Eq. (\ref{eq:bluepot}), it is given by
\be
	\ddot{\varphi} + 3H_{\rm inf} \dot{\varphi} + c H^2_{\rm inf} (\varphi -f) = 0, 
\ee
and we  have
\be
	\varphi  \propto e^{-\lambda N}
	~~~\text{with}~~~ \lambda = \frac{3}{2} -\frac{3}{2} \sqrt{1-\frac{4}{9}c}.
	\label{sol_varphi}
\ee
Since we assume that the initial value of $\varphi$ is far displaced from the minimum, 
the solution (\ref{sol_varphi}) leads to $\varphi_*(k) \propto k^{-\lambda}$ 
for $\varphi \gg f$.
This means that the spectral index of ${\cal P}_{\zeta,{\rm curv}}$, $n_\sigma$, is given by
\be
	n_\sigma = 4 - 3\sqrt{1-\frac{4}{9}c},
\ee
which implies the extremely blue-tilted spectrum up to $n_\sigma = 4$ for $c \sim {9/4}$.
Note that 
the fluctuations of the curvaton which
exit the horizon after $\varphi$ reaches the minimum
is investigated in a similar way to the simple quadratic curvaton model as discussed in
the previous subsection.
As a summary, we can express the power spectrum approximately as
\be
	\mathcal{P}_{\zeta,{\rm curv}}(k) = 
	\begin{cases}
		\mathcal{P}_{\zeta,{\rm curv}}(k_c) \left({k \over k_c }\right)^{n_\sigma-1} ~~~
		&\text{for}~~~ k<k_f \\[1mm]
		\mathcal{P}_{\zeta,{\rm curv}}(k_f)  ~~~
		&\text{for}~~~ k>k_f,
	\end{cases}
	\label{P_zeta_axion}
\ee
where $k_f$ corresponds to the scale exiting the horizon just when $\varphi$ reaches the minimum $f$ and $\mathcal{P}_{\zeta,{\rm curv}}(k_f)$ is calculated as
\be
	\mathcal{P}_{\zeta,{\rm curv}}(k_f) = \mathcal{P}_{\zeta,{\rm curv}}(k_c) \bigg( \frac{k_f}{k_c} \bigg)^{n_\sigma-1}
	= \bigg( \frac{2r(\eta)}{4+3r(\eta)} \bigg)^2 \bigg( \frac{H_{\rm inf}}{2\pi f \theta_i} \bigg)^2.
	\label{P_zeta_k_f}
\ee


\section{Scalar-induced gravitational waves} 
\label{induced_GW}

In this section, we formulate the gravitational waves induced by the scalar metric perturbations and anisotropic stress at second order.

\subsection{Power spectrum of the gravitational waves with a source term}
\label{Sub:evol}

Let us consider the tensor metric perturbations given by
\be
	ds^2 = a^2(\eta) \bigg[ - d\eta^2 + \bigg(\delta_{ij} + \frac{1}{2}h_{ij} \bigg) dx^i dx^j \bigg],
\ee
where 
$h_{ij}$ is the tensor metric perturbation satisfying $h_{ii}=0$ (trace free) and $\partial_ih_{ij}=0$ (transverse).
The time evolution for the metric perturbations is described by the Einstein equation 
and that for $h_{ij}$ is given by
\be
	h''_{ij} + 2 \mathcal{H} h'_{ij} - \nabla^2 h_{ij} = -4 \hat{\mathcal{T}}_{ij}^{~lm} \mathcal{S}_{lm},
	\label{evolve_tensor}
\ee
where the prime represents the derivative with respect to the conformal time, 
$\mathcal{H}$ is defined as $\mathcal{H} = a'/a$, 
$\mathcal{\hat T}_{ij}^{~lm}$ is the projection tensor 
which projects any tensor into the transverse trace-free one, 
and $\mathcal{S}_{ij}$ is the scalar induced source term which is 
given later.
Conventionally, the Fourier transformation of the tensor metric perturbation 
is defined through
\be
	h_{ij}({\bf x},\eta) = \int \frac{d^3k}{(2\pi)^{3/2}} e^{i{\bf kx}} \big[ h_{\bf k}(\eta) e_{ij}({\bf k}) + \bar{h}_{\bf k}(\eta) \bar{e}_{ij}({\bf k}) \big],
\ee
where $e_{ij}({\bf k})$ and $\bar{e}_{ij}({\bf k})$ are two polarization tensors 
defined in terms of two orthonormal basis vectors, $e_i({\bf k})$ and $\bar{e}_i({\bf k})$, 
orthogonal to ${\bf k}$ as
\begin{align}
	e_{ij}({\bf k}) &= \frac{1}{\sqrt{2}} \big[ e_i({\bf k})e_j({\bf k}) - \bar{e}_i({\bf k})\bar{e}_j({\bf k}) \big], \\[1mm]
	\bar{e}_{ij}({\bf k}) &= \frac{1}{\sqrt{2}} \big[ e_i({\bf k})\bar{e}_j({\bf k}) + \bar{e}_i({\bf k}) e_j({\bf k}) \big].
\end{align}
Using these polarization tensors, the projection tensor $\hat{\mathcal{T}}_{ij}^{~lm}$ 
is defined via
\be
	\hat{\mathcal{T}}_{ij}^{~lm} \mathcal{S}_{lm}= \int \frac{d^3k}{(2 \pi)^{3/2}} e^{i{\bf kx}} 
	\big[ e_{ij}({\bf k}) e^{lm}({\bf k}) + \bar{e}_{ij}({\bf k}) \bar{e}^{lm}({\bf k}) \big] \mathcal{S}_{lm}({\bf k},\eta)
\ee
where $\mathcal{S}_{ij}({\bf k},\eta)$ is the Fourier transformed component of 
$\mathcal{S}_{ij}({\bf x},\eta)$.
Thus the evolution equation for the Fourier transformed component of $h_{ij}$ is derived as
\be
	h''_{\bf k} + 2 \mathcal{H} h'_{\bf k} + k^2 h_{\bf k} = \mathcal{S}({\bf k},\eta),
	\label{evolve}
\ee
where $\mathcal{S}({\bf k},\eta) = -4 e^{lm}\mathcal{S}_{lm}({\bf k},\eta)$.
Now we solve Eq. (\ref{evolve}) by using the Green's function method.
The solution is found to be 	
\be
	h_{\bf k}(\eta) = \frac{1}{a(\eta)} \int^{\eta}_{\eta_0} d\tilde{\eta} a(\tilde\eta) g_{\bf k}(\eta;\tilde\eta) \mathcal{S}({\bf k},\tilde\eta),
	\label{solution}
\ee
where $g_{\bf k}(\eta,\tilde\eta)$ is the Green's function which is defined through
\be
	g_{\bf k}''(\eta; \tilde \eta) + \bigg(k^2 - \frac{a''(\eta)}{a(\eta)} \bigg) g_{\bf k}(\eta;\tilde\eta) = \delta(\eta - \tilde\eta).
\ee
In particular, the Green's function is given by
\be
	g_{\bf k}(\eta;\tilde\eta) = \frac{\sin[ k(\eta-\tilde\eta)]}{k} \theta(\eta-\tilde\eta)
\ee
in radiation dominated era and
\be
	g_{\bf k}(\eta; \tilde\eta) = \frac{(k^2\eta \tilde\eta + 1)\sin[k(\eta-\tilde\eta)] - k(\eta- \tilde\eta) \cos[k(\eta-\tilde\eta)]}{k^3\eta \tilde\eta} \theta(\eta-\tilde\eta)
\ee
in matter dominated era.
Then, the power spectrum is formally given by
\begin{align}
	\langle h_{\bf  k}(\eta) h_{\bf p}(\eta) \rangle &= 
	{1 \over a(\eta)^2} \int^\eta_{\eta_0}d\tilde{\eta}_1 a(\tilde{\eta}_1)
	g_{\bf k}(\eta; \tilde{\eta}_1)\int^\eta_{\eta_0} d \tilde{\eta}_2 a(\tilde{\eta}_2)
	g_{\bf p}(\eta; \tilde{\eta}_2) \langle \mathcal{S}({\bf k}, \tilde{\eta}_1) \mathcal{S}({\bf p}, \tilde{\eta}_2)\rangle
	\nonumber \\[1mm]
	&\equiv
	\frac{2 \pi^2}{k^3} \delta^3({\bf k}+{\bf p}) \mathcal{P}_h(k,\eta) ,
	\label{P_h}
\end{align}
and the energy density of gravitational waves is given by \cite{Maggiore:1999vm}
\be
	\rho_{\rm GW}(\eta) 
	= \int d\ln k~ \frac{k^2}{16 \pi G a^2}  \mathcal{P}_h(k,\eta).
\ee
In order to discuss the amount of the present gravitational wave background at the present,
it is useful to introduce
a density parameter of gravitational waves within the logarithmic interval of wavenumber which is given by
\be
	\Omega_{\rm GW} (k,\eta) = \frac{1}{\rho_{\rm cr}(\eta)} \frac{d \rho_{\rm GW}(\eta)}{d \ln k} = \frac{k^2}{6\mathcal{H}^2(\eta)} \mathcal{P}_h(k,\eta), 
\ee
where $\rho_{\rm cr}$ is the critical density.
Focusing only on those gravitational waves generated before the matter radiation equality, $\Omega_{\rm GW}$ today is calculated via
\be
	\Omega_{\rm GW} (k) = \frac{k^2 \Omega_\gamma }{6\mathcal{H}^2(\eta_\star)} \mathcal{P}_h(k,\eta_\star), 
	\label{Eq;Omega_GW}
\ee
where $\Omega_\gamma \simeq 4.8 \times 10^{-5}$ is the density parameter of radiation today,
and a subscript $\star$ denotes a certain time 
after the amplitude of the gravitational wave, $|h_{\bf k}|$, starts monotonically to decrease as $1/a$
on sub-horizon scales during radiation dominated era.

\subsection{Source term $\mathcal{S}_{ij}$}

In the first order perturbation theory, the right hand side of (\ref{evolve_tensor}), 
$\mathcal{S}_{ij}$, is zero 
when the anisotropic stress does not appear in that order.
However, 
once the second order contributions are included, 
it induces non-zero $\mathcal{S}_{ij}$.
In the curvaton scenario considered here, we have two contributions to the source term at the second order;
one is coming from the scalar adiabatic curvature perturbations and another is
the anisotropic stress due to the kinetic term of the curvaton before the curvaton decay.
In this subsection, we discuss these two sources separately.

\subsubsection{Second order adiabatic curvature perturbations}

\begin{figure}[tp]
\centering
\includegraphics [width = 8cm, clip]{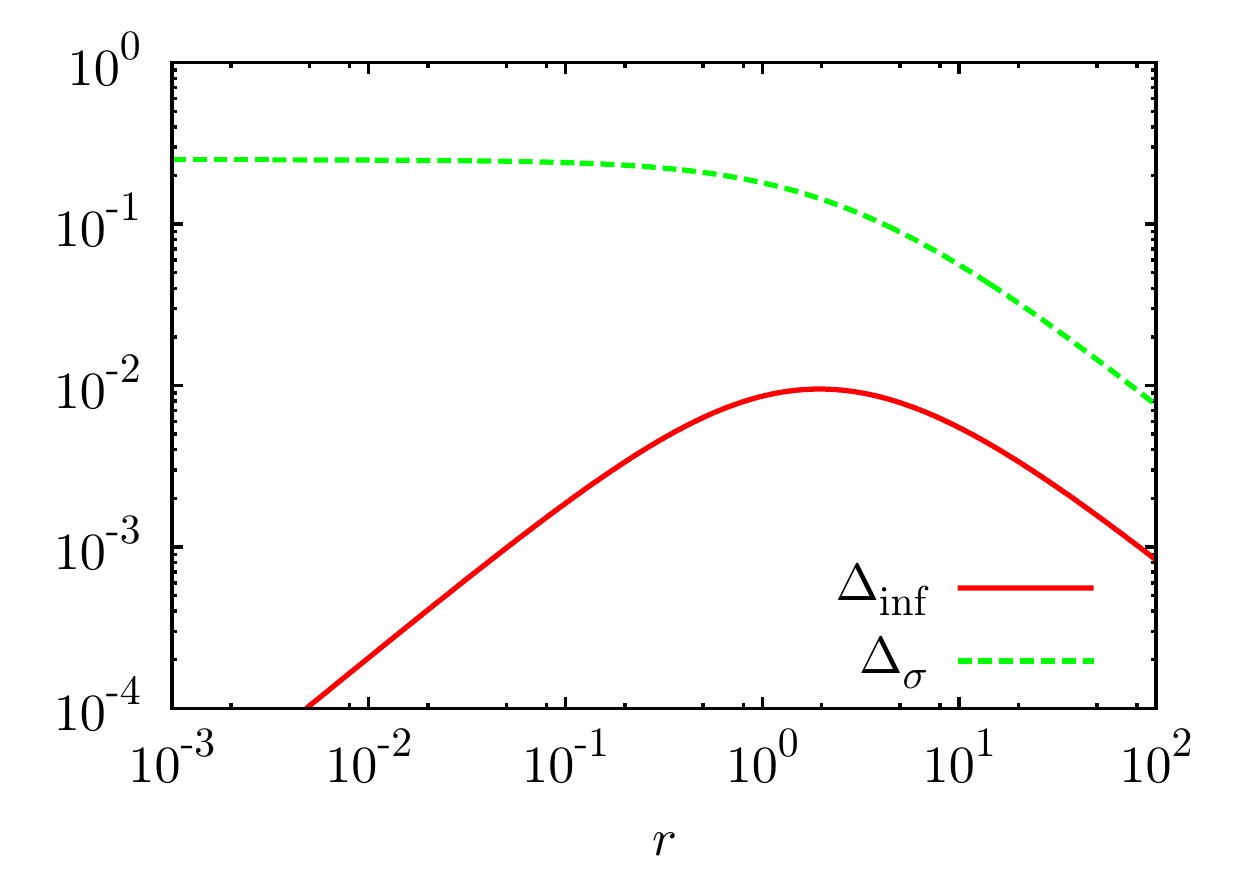}
\caption{
	The deviation from $\Phi' = 0$ defined via (\ref{deviation}) is shown.
	The horizontal axis shows $r(\eta)$.
	The solid red line and dashed green line correspond to the inflaton and the curvaton contributions respectively.
}
\label{Fig;Phi}
\end{figure}

Following the literature, let us consider the scalar metric perturbations in conformal Newtonian gauge
which is described as
\be
	ds^2 = a^2(\eta) \bigg[ -(1+2\Phi) d\eta^2 + (1-2\Psi)\delta_{ij}  dx^i dx^j \bigg].
\ee
At the second order, the source term induced from the above scalar metric perturbations
can be written as ~\cite{Acquaviva:2002ud}
\be
	\mathcal{S}_{ij}^{\Phi} = 
	- 2\partial_i\Phi \partial_j\Phi 
	- \mathcal{H}^{-2} \partial_i(\Phi' + \mathcal{H}\Phi) \partial_j(\Phi'+\mathcal{H}\Phi) ,
	\label{eq:sphi}
\ee
where 
we 
have assumed  that 
the first order anisotropic stress is zero and it leads to $\Phi = \Psi$.
Then, $\mathcal{S}^{\Phi }({\bf k})$, which appears in the right hand side of the evolution equation
for the Fourier transformed component of $h_{ij}$,
is calculated through the convolution 
of two Fourier modes as
\be
	\begin{split}
		\mathcal{S}^{\Phi}({\bf k},\eta) &=  -4e^{lm}({\bf k}) \mathcal{S}_{lm}^{\Phi}({\bf k},\eta) \\[1mm]
		&=  4 \int\frac{d^3\tilde{k}}{(2 \pi)^{3/2}} e^{lm}({\bf k}) \tilde{k}_l \tilde{k}_m 
		\bigg[ 3\Phi_{\bf \tilde{k}}(\eta) \Phi_{\bf k-\tilde{k}}(\eta) \\[1mm]
		&~~~~~~~~~~~~~~~~~~~~~~~~~~~~~~~~~~~
		 +\frac{2}{\mathcal{H}} \Phi'_{\bf \tilde{k}}(\eta) \Phi_{\bf k-\tilde{k}}(\eta) 
		+ \frac{1}{\mathcal{H}^2} \Phi'_{\bf \tilde{k}}(\eta) \Phi'_{\bf k-\tilde{k}}(\eta) 
		 \bigg].
	\end{split}
	\label{source_term_Phi}
\ee
Let us consider the evolution of $\Phi$.
The gravitational potential in conformal Newtonian gauge, $\Phi$,
is related to  
the curvature perturbation on uniform density slicing, 
considered in the previous section, as~\cite{Kodama:1985bj,Mukhanov:1990me}
\be
	-\zeta = \Psi + \frac{H}{\dot\rho} \delta\rho = \frac{6+5r}{4+3r} \Phi + \frac{2+2r}{4+3r} \frac{\Phi'}{\mathcal H},
\ee
where $r$ is the ratio of the energy density of the curvaton to that of radiation defined in
the previous section.
Note that in the curvaton mechanism $\Phi' \neq 0$ before the curvaton decay 
even on the super-horizon scales. 
Denoting $\zeta = \zeta_{\rm inf} + (r(\eta)/(4+3r(\eta)))S_\sigma$ and following the above relation,
we obtain the evolution equation for $\Phi$ on superhorizon scales as
\be
	\frac{\Phi'(\eta)}{\mathcal H(\eta)} + \frac{6+5r(\eta)}{2+2r(\eta)} \Phi(\eta) + \frac{4+3r(\eta)}{2+2r(\eta)}\zeta_{\rm inf} + \frac{r(\eta)}{2+2r(\eta)} S_\sigma = 0.
	\label{evol_Phi}
\ee
This equation can be analytically solved as shown in appendix \ref{app:grav_pot} and the solution is formally given by
\be
	\Phi(\eta) = -\frac{4+3r(\eta)}{6+5r(\eta)} \left(1+\Delta_{\rm inf}(\eta)\right) \zeta_{\rm inf} - \frac{r(\eta)}{6+5r(\eta)} \left(1-\Delta_\sigma(\eta) \right)S_\sigma,
	~~~\text{(superhorizon)}
	\label{deviation}
\ee
where $\Delta_{\rm inf}$ and $\Delta_\sigma$ represent the deviations from $\Phi' = 0$ and are shown in Fig.~\ref{Fig;Phi}.
This figure shows that $\Phi$ deviates from the value with $\Phi' = 0$ at most $25\%$.
Although the deviations are not so large,
in the following numerical calculation of the scalar-induced gravitational wave background,
we properly include the contribution from $\Phi'$ on super-horizon scales.
%
%
%
For subhorizon modes, $\Phi$ is constrained by the Poisson equation and we obtain
\be
	\Phi(\eta) = -\frac{3}{2} \bigg( \frac{\mathcal{H}(\eta)}{k} \bigg)^2 \bigg( \frac{1}{1+r(\eta)} \frac{\delta\rho_r}{\rho_r} + \frac{r(\eta)}{1+r(\eta)} \frac{\delta\rho_\sigma}{\rho_\sigma} \bigg).
	~~\text{(subhorizon)}
	\label{subhorizon}
\ee
Note that $\Phi$ is frozen after the curvaton starts to dominate the universe until its decay if $r_D > 1$.
Focusing only on the curvaton contribution, we unify (\ref{deviation}) and (\ref{subhorizon}), which can be expressed as
\be
	\Phi(\eta (< \eta_{\rm dec})) \approx 
	\begin{cases}
		T_S(k,\eta) S_\sigma &\text{for}~~~\eta < \eta_{\rm dom} \\[1mm]
		T_S(k,\eta_{\rm dom} ) S_\sigma &\text{for}~~~\eta > \eta_{\rm dom},
	\end{cases}
\ee
where $\eta_{\rm dec}$ and $\eta_{\rm dom}$ are the conformal time at the decay and at the beginning of curvaton domination respectively and we define
\be
	T_S(k,\eta) = -\frac{r(\eta)}{6+5r(\eta)}(1-\Delta_\sigma) S_\sigma \bigg[ 1+\frac{2(1+r(\eta))}{3(6+5r(\eta))} (1-\Delta_\sigma)(k\eta)^2 \bigg]^{-1}.
\ee
The transfer function of $\Phi$ after the curvaton decay in radiation dominated era is given by a decaying-oscillation solution as \cite{Dodelson:1282338}
\be
	T_\Phi(k,\eta) = \frac{3}{(k\eta/\sqrt{3})^3} 
	\bigg[ \sin\bigg( \frac{k\eta}{\sqrt{3}} \bigg) - \frac{k\eta}{\sqrt{3}} \cos\bigg(\frac{k\eta}{\sqrt{3}}\bigg) \bigg],
	\label{transfer_Phi}
\ee
and we obtain the time evolution of $\Phi$ after the curvaton decay as
\be
	\Phi(k,\eta(>\eta_{\rm dec})) \approx 
	\begin{cases} 
		T_\Phi(k,\eta) T_S(k,\eta_{\rm dec}) S_\sigma ~~~&\text{for}~~~k < 1/\eta_{\rm dec} \\[1mm]
		T_\Phi(1/\eta_{\rm dec},\eta) T_S(k,\eta_X) S_\sigma ~~~&\text{for}~~~k > 1/\eta_{\rm dec},
	\end{cases}
\ee
where we define $\eta_X = \eta_{\rm dec}$ for $r_D < 1$ and $\eta_X = \eta_{\rm dom}$ otherwise.

Assuming that there exists the curvaton dominated epoch, $r_D > 1$, the resultant density parameter of gravitational waves is enhanced on scales which are inside the horizon 
in curvaton dominated era because $\Phi$ remains constant at that time.
This leads to $h_{\bf k} \simeq {\rm const.}$ on subhorizon scales \cite{Assadullahi:2009nf,Alabidi:2013lya}.
On the other hand, gravitational waves emitted before the curvaton domination, 
the energy density of gravitational waves suffers an additional suppression because of $\rho_{\rm GW}/\rho_\sigma \propto a^{-1}$.
Thus, the present energy density spectrum of gravitational waves (\ref{Eq;Omega_GW}) is expressed as
\be
	\Omega_{\rm GW} (k) = \frac{k^2 \Omega_\gamma }{6\mathcal{H}^2(\eta_k)} F^2 \mathcal{P}_h(k,\eta_k), 
\ee
where the subscript $k$ denotes the value at the horizon reentering and $F^2$ denotes the enhancement 
and suppression due to the early matter (curvaton) domination, which is given by
\be
	F^2 \approx \begin{cases}
		(k\eta_{\rm dec})^2 &\text{for}~~1/\eta_{\rm dec}< k < 1/\eta_{\rm dom}, \\[1mm]
		(k\eta_{\rm dom})^{-4}(\eta_{\rm dec}/\eta_{\rm dom})^2 + (\eta_{\rm dom}/\eta_{\rm dec})^2 &\text{for}~~1/\eta_{\rm dom}< k \\[1mm]
		1 &\text{otherwise}.
	\end{cases}
	\label{F^2}
\ee
Here we assume the density perturbation on subhorizon scales, which grows proportional to the scale factor in matter dominated universe, does not reach $\mathcal{O}(1)$.
Otherwise, the enhancement ends at the cut off scale $k_{\rm NL} \sim \mathcal{P}^{-1/4}_{\zeta,{\rm curv}} (1/\eta_{\rm dec}) /\eta_{\rm dec}$ 
corresponding to the scale which becomes nonlinear at the curvaton decay and $1/\eta_{\rm dom}$ in (\ref{F^2}) must be replaced with $k_{\rm NL}$.

\subsubsection{Anisotropic stress sourced from the kinetic term of the curvaton} 
 
Next, in the curvaton scenario, there is another source induced from the kinetic term of the curvaton as
\be
	\mathcal{S}^{\rm kin}_{ij} =  M_P^{-2} \partial_i\delta\sigma\partial_j\delta\sigma,
	\label{eq:skin}
\ee
and the source term of the evolution equation for $h_{\rm k}$ is given by
\begin{align}
		\mathcal{S}^{\rm kin}({\bf k},\eta) &=  -4e^{lm}({\bf k}) \mathcal{S}_{lm}^{\rm kin}({\bf k},\eta) 
		\nonumber\\[1mm]
		&=  -4 M_P^{-2} \int\frac{d^3\tilde{k}}{(2 \pi)^{3/2}} e^{lm}({\bf k}) \tilde{k}_l \tilde{k}_m 
		\delta\sigma_{\bf \tilde{k}}(\eta) \delta\sigma_{\bf k-\tilde{k}}(\eta).
	\label{source_term_kin}
\end{align}
By decomposing $\delta \sigma$ into the transfer function and the initial value evaluated on super-horizon scales
at a certain time before the curvaton starts to oscillate as
\be
     \delta\sigma_{\bf k}(\eta) = T_\sigma (k,\eta) \delta\sigma_{\bf k},
\ee
the transfer function, $T_\sigma (k,\eta)$,
 is derived from the evolution equation for $\delta\sigma$ for each Fourier mode.
We summarize the behavior of $\delta\sigma$ below.
\begin{itemize}
	\item
	For Fourier modes reentering the horizon after the curvaton starts to oscillate at $\eta_{\rm osc}$, 
	$\delta\sigma$ remains frozen for $\eta < \eta_{\rm osc}$ on super-horizon scales
	and evolves like $a^{-3/2}$ for $\eta > \eta_{\rm osc}$ both on super-horizon and sub-horizon scales.
	Hence we obtain the transfer function for $k\eta_{\rm osc} < 1$ as
	\be
		T_\sigma\left(k (< 1/ \eta_{\rm osc}),\eta\right) = 
		\begin{cases}
		1 ~~~&\text{for}~~~\eta < \eta_{\rm osc} \\
		\big( \frac{\eta_{\rm osc}}{\eta} \big)^{3/2} ~~~&\text{for}~~~ \eta > \eta_{\rm osc}.
		\end{cases} 
		\label{transfer_sigma_1}
	\ee
	\item
	Oppositely, for modes reentering the horizon before the curvaton oscillation, $\delta\sigma$ remains constant for $\eta < 1/k$, 
	evolves like $\propto a^{-1}$ for $1/k < \eta < \eta_m$ and $\propto a^{-3/2}$ for $\eta > \eta_m$, where $\eta_m$ is defined via $k/a(\eta_m) = m_\sigma$ 
	and given by $\eta_m = k\eta_{\rm osc}^2$ \cite{Bartolo:2007vp}.
	Thus we obtain the transfer function for $k\eta_{\rm osc} > 1$ as
	\be
		T_\sigma \left(k (> 1 / \eta_{\rm osc}),\eta \right) = 
		\begin{cases} 
		1 ~~~&\text{for}~~~\eta < 1/k \\
		\frac{1}{k\eta} ~~~&\text{for}~~~1/k < \eta < \eta_m \\
		\frac{1}{k\eta_m} \big( \frac{\eta_m}{\eta} \big)^{3/2} ~~~&\text{for}~~~\eta > \eta_m.
		\end{cases}
		\label{transfer_sigma_2}
	\ee
\end{itemize}

\section{Results} \label{sec;results}

Based on the above formulation, let us evaluate the amount of the gravitational waves induced from
the scalar fluctuations in the curvaton scenario.

\subsection{Contribution from $\mathcal{S}_{ij}^{\Phi}$}

Let us consider the contribution from $\mathcal{S}_{ij}^{\Phi}$ given by Eq.~(\ref{eq:sphi}).
This has been well investigated in the literature
and the resultant density parameter for gravitational waves can be approximated as \cite{Ananda:2006af,Baumann:2007zm}
\be
	\Omega_{\rm GW} \sim 10^{-19} \bigg( \frac{\mathcal{P}_{\zeta,{\rm curv}}(k)}{\mathcal{P}_\zeta(k_c)} \bigg)^2, 
	\label{approx_Omega}
\ee
which is also confirmed by our full numerical integration shown in appendix \ref{app:power}.
This result is quite reasonable since the gravitational waves are sourced by the quadratic of the curvature perturbation.
In the curvaton model with blue spectrum, $\Omega_{\rm GW}$ has a peak at the wave number $k_{\rm dec}$ given by
\be
	k_{\rm dec} \simeq 1.7 \times 10^{15}~{\rm Mpc}^{-1} \bigg( \frac{\Gamma_\sigma}{1~{\rm GeV}} \bigg)^{1/2},
	\label{eq:k_dec}
\ee
for $r_D < 1$ because the power spectrum of curvature perturbations from the curvaton is suppressed 
proportional to $r=\rho_\sigma/\rho_r \propto a$ before the curvaton decay.
For $r_D > 1$, the peak wave number is either $k_{\rm dom} = 1/\eta_{\rm dom}$ or $k_{\rm NL}$ given by
\be
	k_{\rm peak} = {\rm min}(k_{\rm dom},k_{\rm NL}) \sim {\rm min}(r_D^{1/2} k_{\rm dec}, \mathcal{P}_{\zeta,{\rm curv}}(k_{\rm dec})^{-1/4} k_{\rm dec}).
\ee
Taking into account the above, we show the resultant spectrum of the gravitational waves in the quadratic curvaton model and the axion-like curvaton model below.

\subsubsection{Quadratic curvaton model}

\begin{figure}[tp]
\centering
\subfigure[]{
\includegraphics [width = 7.5cm, clip]{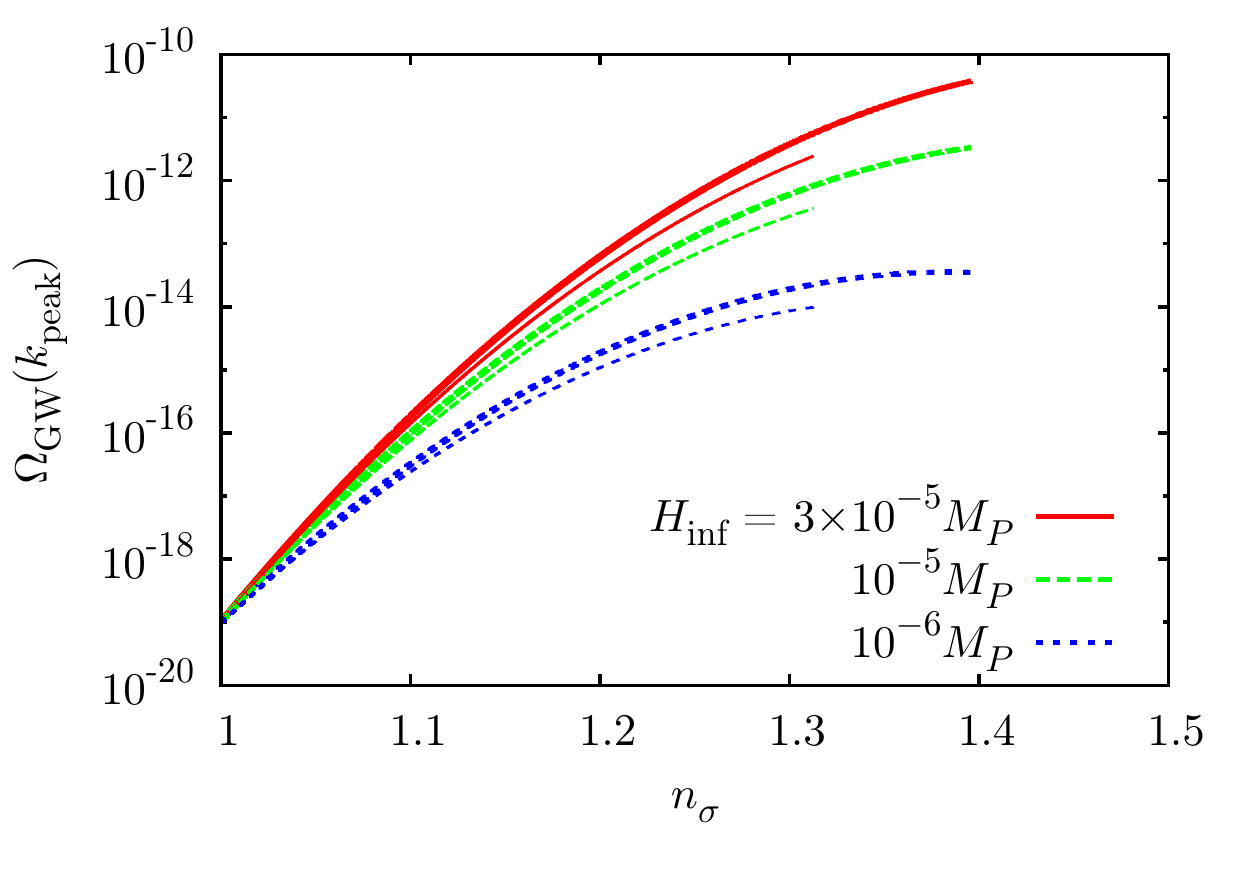}
\label{n_sigma_q_a}
}
\subfigure[]{
\includegraphics [width = 7.5cm, clip]{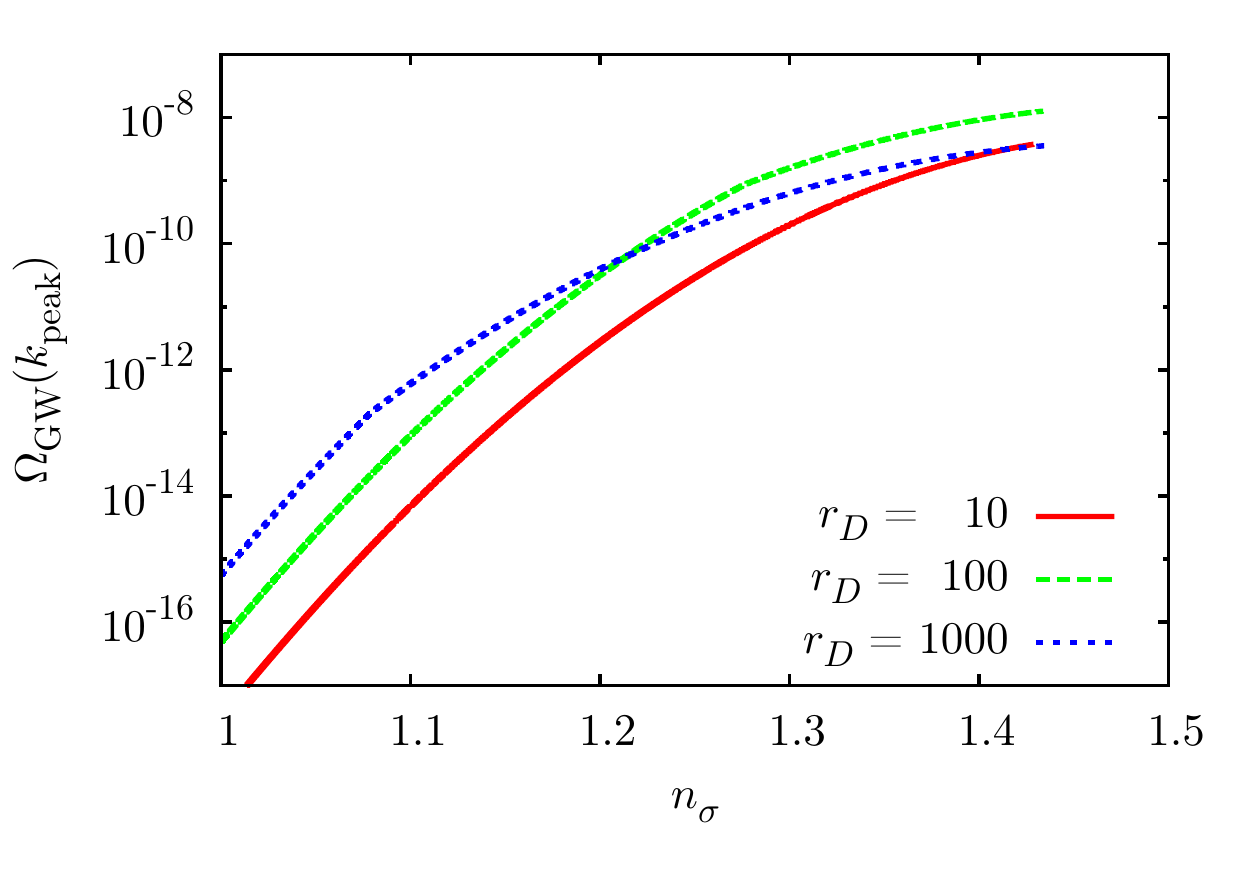}
\label{n_sigma_q_b}
}
\caption{
	The peak values of $\Omega_{\rm GW}$ in quadratic curvaton model are shown.
	The horizontal axises is the spectral index $n_\sigma$.
	We have taken $H_{\rm inf} = 3 \times 10^{-5}M_P$ (solid red), $10^{-5}M_P$ (dashed green) and $10^{-6}M_P$ (dotted blue), 
	$r_D = 1$ (thick lines) and $r_D = 0.1$ (thin lines) in the left panel 
	and $H_{\rm inf} = 3 \times 10^{-5}M_P$, $r_D = 10$ (solid red), $r_D = 100$ (dashed green) and $r_D = 1000$ (dotted blue) in the right panel.
}
\label{n_sigma_q}
\end{figure}

\begin{figure}[tp]
\centering
\subfigure[]{
\includegraphics [width = 7.5cm, clip]{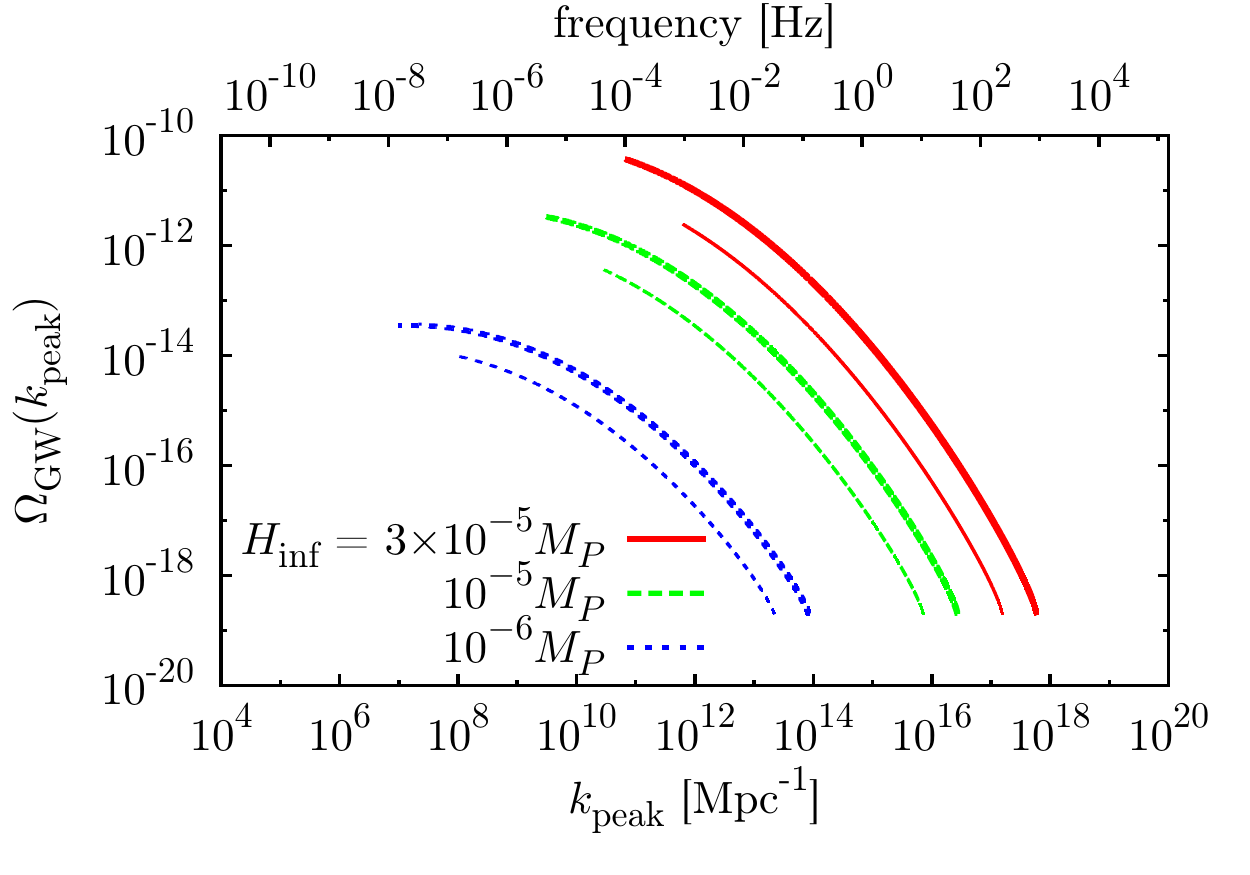}
\label{k_dec_q_a}
}
\subfigure[]{
\includegraphics [width = 7.5cm, clip]{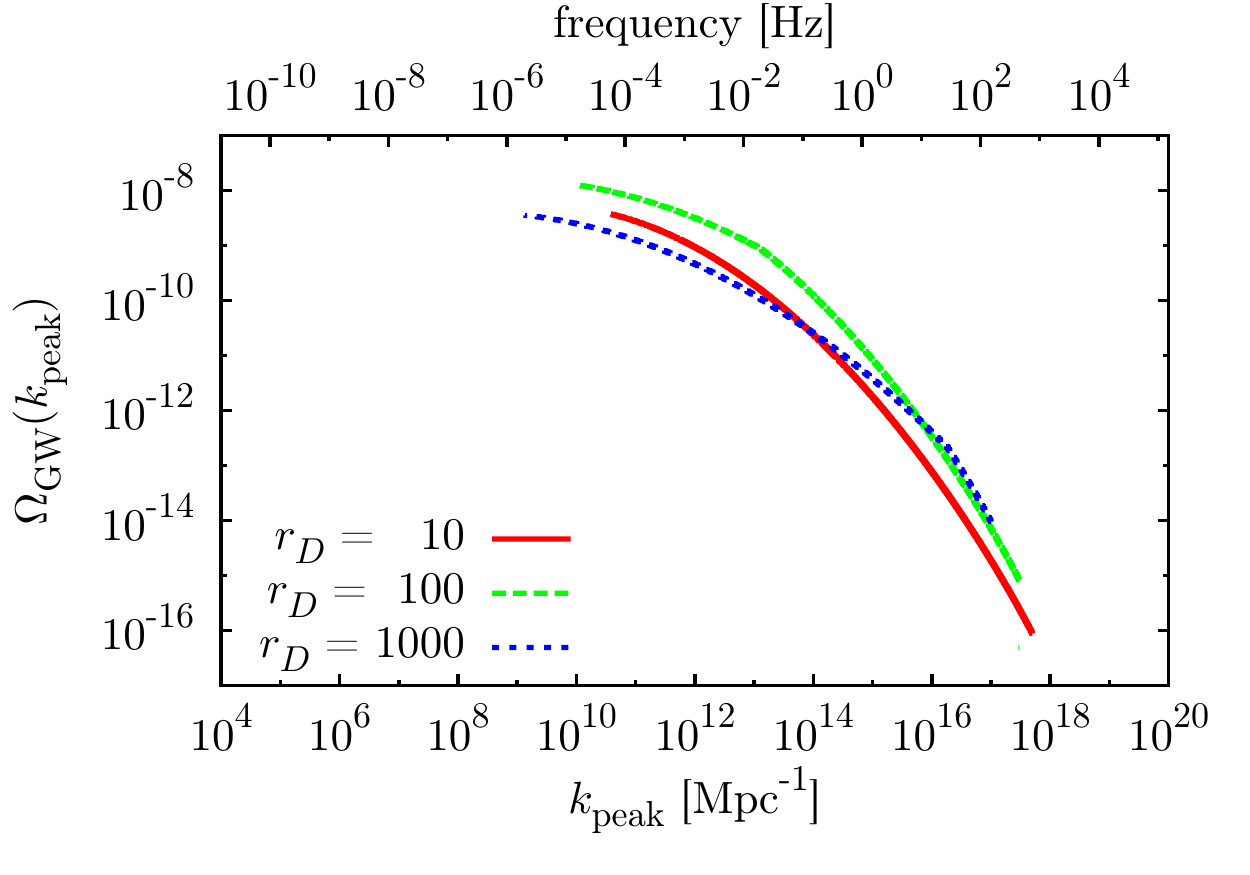}
\label{k_dec_q_b}
}
\caption{
	The peak values of $\Omega_{\rm GW}$ in quadratic curvaton model are shown.
	The horizontal axises is the wave number at peak $k_{\rm peak}$ or corresponding frequency.
	We have taken $H_{\rm inf} = 3 \times 10^{-5}M_P$ (solid red), $10^{-5}M_P$ (dashed green) and $10^{-6}M_P$ (dotted blue), 
	$r_D = 1$ (thick lines) and $r_D = 0.1$ (thin lines) in the left panel 
	and $H_{\rm inf} = 3 \times 10^{-5}M_P$, $r_D = 10$ (solid red), $r_D = 100$ (dashed green) and $r_D = 1000$ (dotted blue) in the right panel.
}
\label{k_dec_q}
\end{figure}

\begin{figure}[tp]
\centering
\subfigure[]{
\includegraphics [width = 7.5cm, clip]{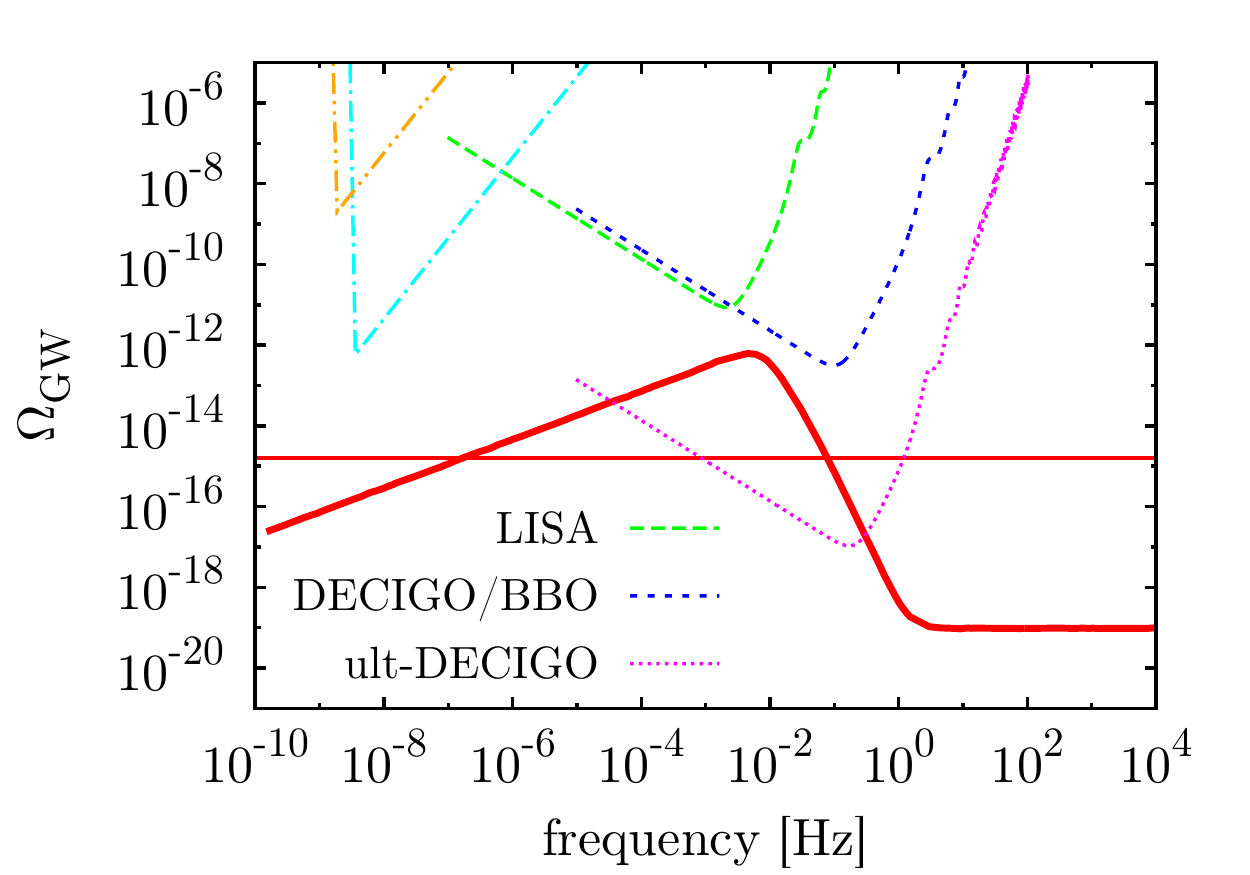}
\label{Omega_q_1}
}
\subfigure[]{
\includegraphics [width = 7.5cm, clip]{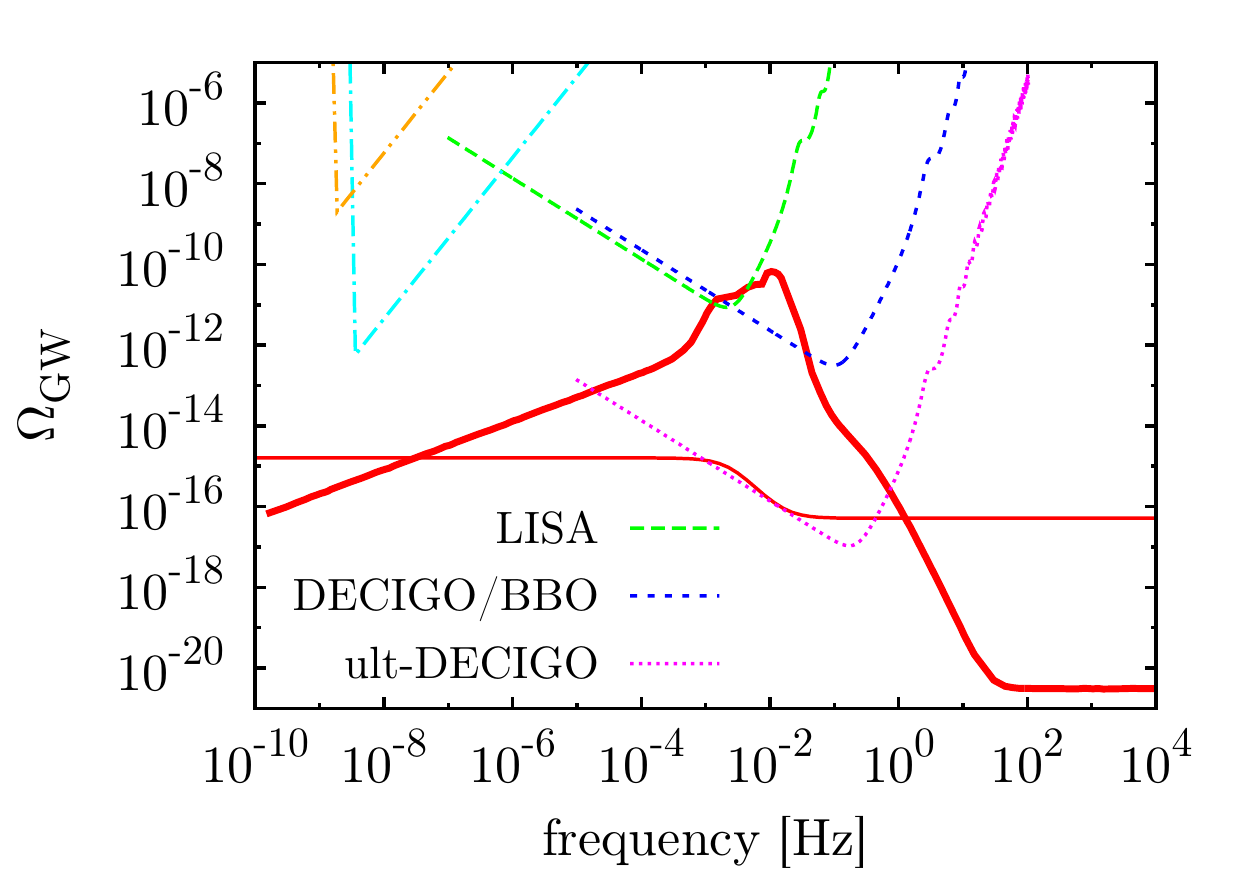}
\label{Omega_q_2}
}
\caption{
	The spectrum of $\Omega_{\rm GW}$ in the quadratic curvaton model is shown as the thick solid red line.
	We have taken $H_{\rm inf} = 3 \times 10^{-5}M_P$, $n_\sigma = 1.3$, $r_D = 1$ (left panel) and $r_D = 10$ (right panel).
	The thin solid red line represents the contribution from the primordial tensor metric perturbation.
	We also show the sensitivity curves of LISA (dashed green), DECIGO/BBO (dotted blue), ultimate-DECIGO (small dotted magenta) 
	and the pulsar timing observation by SKA (dash dotted cyan).
	The dash double-dotted orange line correspond to the current upper limit from the pulsar timing.
}
\label{Omega_q}
\end{figure}

First, we consider the induced gravitational waves in the quadratic curvaton model.
Here we assume $m_\sigma \simeq \Gamma_\phi$ which means that the reheating occurs soon after inflation, 
because this gives the maximal amount of gravitational waves once we fix $r_D$, $H_{\rm inf}$ and $n_\sigma$.
We show the peak value of $\Omega_{\rm GW}$ written as $\Omega_{\rm GW}(k_{\rm peak})$ 
in terms of the spectral index $n_\sigma$ in Fig.~\ref{n_sigma_q} and the wave number at the peak $k_{\rm peak}$ or corresponding frequency in Fig.~\ref{k_dec_q}.
In these figures, we have used the approximation (\ref{approx_Omega}) 
and taken $H_{\rm inf} = 3 \times 10^{-5}M_P$ (red solid), $10^{-5}M_P$ (dashed green) and $10^{-6}M_P$ (dotted blue), 
$r_D = 1$ (thick lines) and $r_D = 0.1$ (thin lines) in the left panels and $H_{\rm inf} = 3 \times 10^{-5}M_P$, $r_D = 10$ (solid red), $r_D = 100$ (dashed green) and $r_D = 1000$ (dotted blue) in the right panels.
Upper limit of these lines correspond to the breakdown of $\sigma > H_{\rm inf}/2 \pi$.
We have found that the maximal amount of gravitational waves is about $\Omega_{\rm GW} \sim 10^{-10}$ for $r_D < 1$ and $\Omega_{\rm GW} \sim 10^{-8}$ for $r_D > 1$ 
with a peak frequency $\sim 10^{-4}$~Hz.
Note that the decay rate of the curvaton is proportional to $r_D^{-2}$ from (\ref{r_D_1}), so the decay of the curvaton is delayed for large $r_D$, 
which implies the smaller $k_{\rm NL}$ and the amount of gravitational waves decreases.
We have found that $r_D \sim 100$ maximizes the amount of gravitational waves in this model.
Fig.~\ref{Omega_q} shows the numerical resultant power spectrum of $\Omega_{\rm GW}$ (solid thick red).
In this figure, we have taken $H_{\rm inf} = 3 \times 10^{-5}M_P$, $n_\sigma = 1.3$, $r_D = 1$ (left panel) and $r_D = 10$ (right panel).
The thin line corresponds to the contribution from the primordial tensor metric perturbations and 
sensitivity curves\footnote{
	We have used the online sensitivity curve generator. \url{http://www.srl.caltech.edu/~shane/sensitivity/}
}
of LISA (dashed green), DECIGO/BBO (dotted blue), ultimate-DECIGO (small dotted magenta) 
and the pulsar timing observation by SKA (dash dotted cyan) are also shown.
The region above the dash double-dotted orange line is already ruled out by pulsar timing observations.
We have found that the amount of induced gravitational waves can overcome that of the primordial ones.
Note that the shape of the spectrum varies depending on whether the universe experience the curvaton dominated epoch or not,
 so we can distinguish the model with $r_D < 1$ and that with $r_D > 1$.

\subsubsection{Axion-like curvaton model}

\begin{figure}[tp]
\centering
\subfigure[]{
\includegraphics [width = 7.5cm, clip]{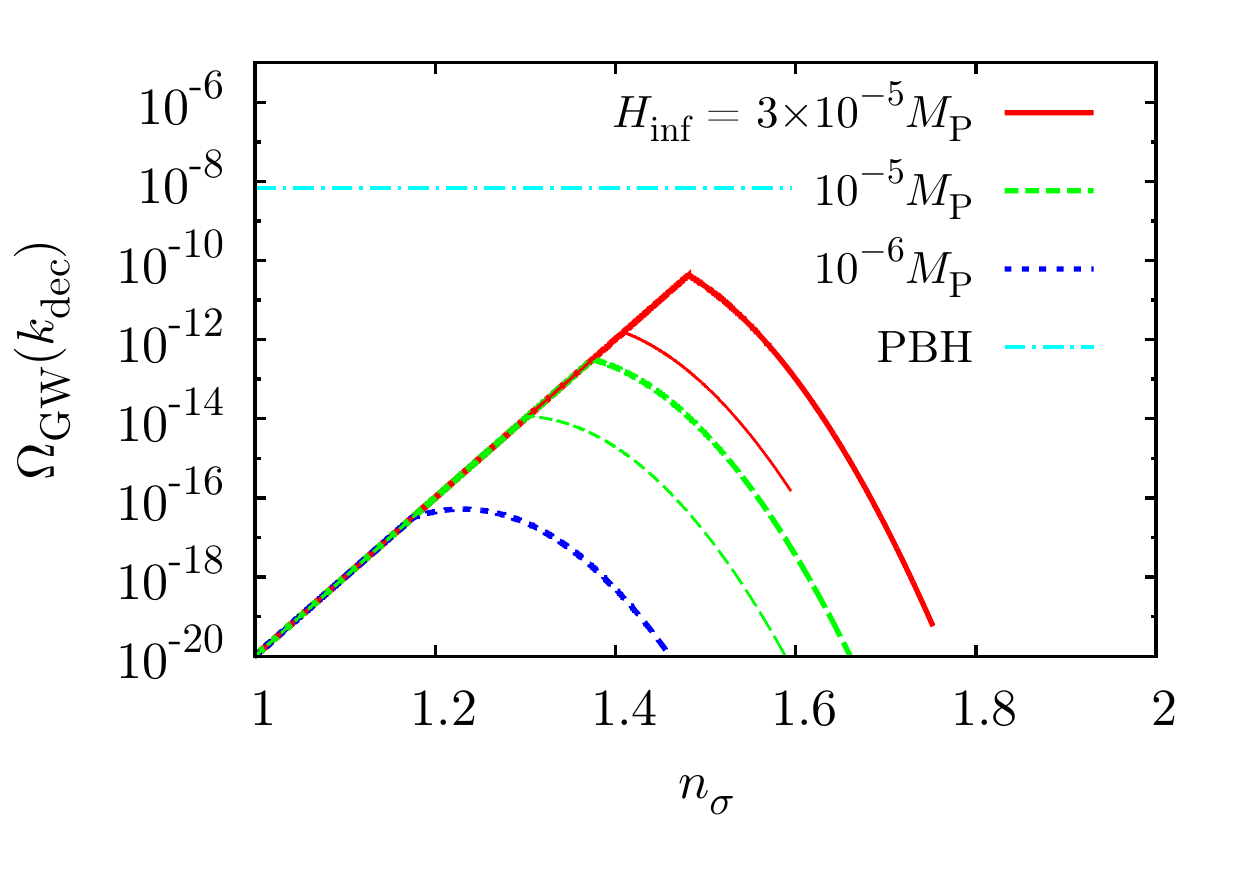}
\label{n_sigma_a_1}
}
\subfigure[]{
\includegraphics [width = 7.5cm, clip]{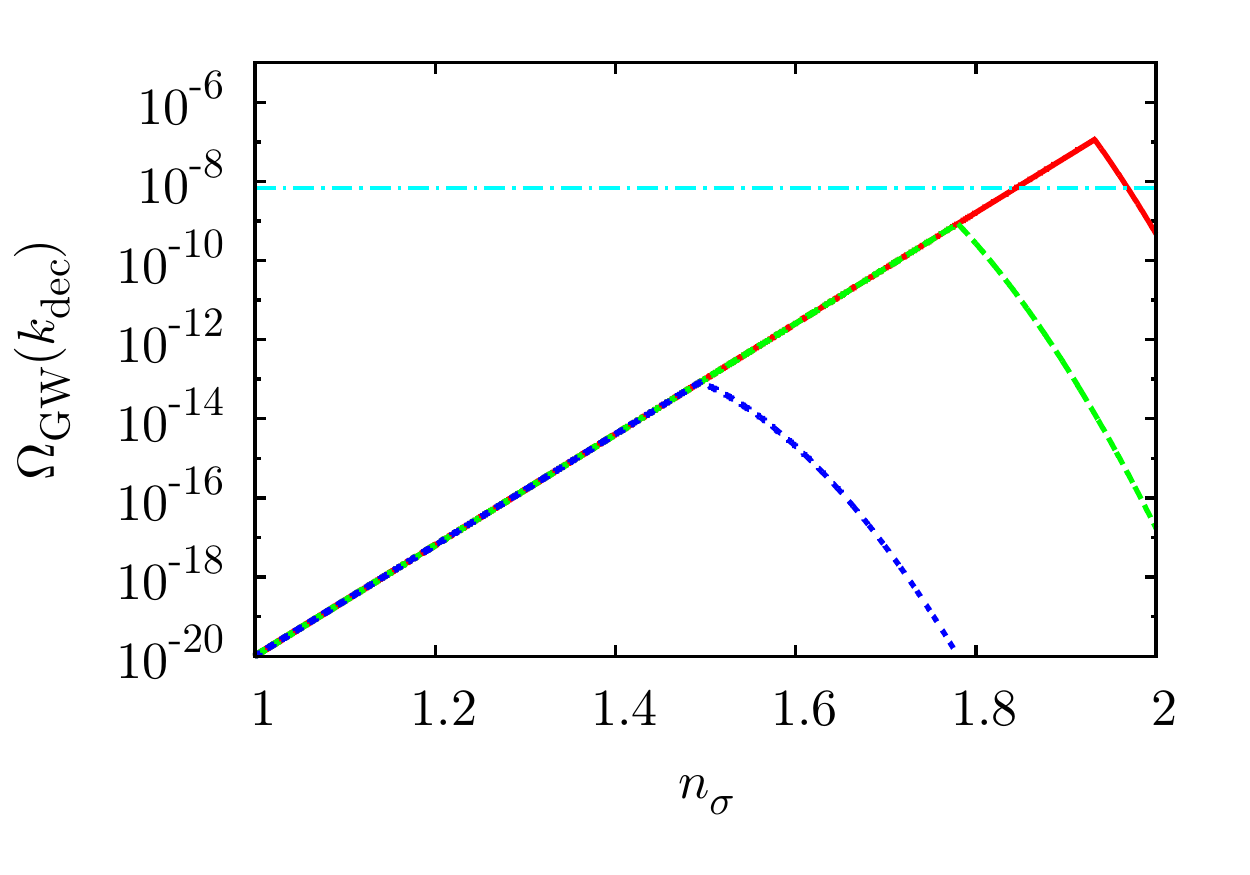}
\label{n_sigma_a_2}
}
\caption{
	The peak values of $\Omega_{\rm GW}$ in terms of the spectral index $n_\sigma$ in axion-like curvaton model are shown.
	We have taken $k_f = 10^{10}~(10^{7})~{\rm Mpc^{-1}}$ for left (right) panel, $\kappa = 1$, 
	$H_{\rm inf} = 3 \times 10^{-5}~M_P$ (solid red), $10^{-5}~M_P$ (dashed green) and $10^{-6}~M_P$ (dotted blue), 
	$r_D = 1$ (thick lines) and $r_D = 0.1$ (thin lines).
	The dash dotted cyan line shows the upper bound from the PBH overproduction.
}
\label{n_sigma_a}
\end{figure}

\begin{figure}[tp]
\centering
\subfigure[]{
\includegraphics [width = 7.5cm, clip]{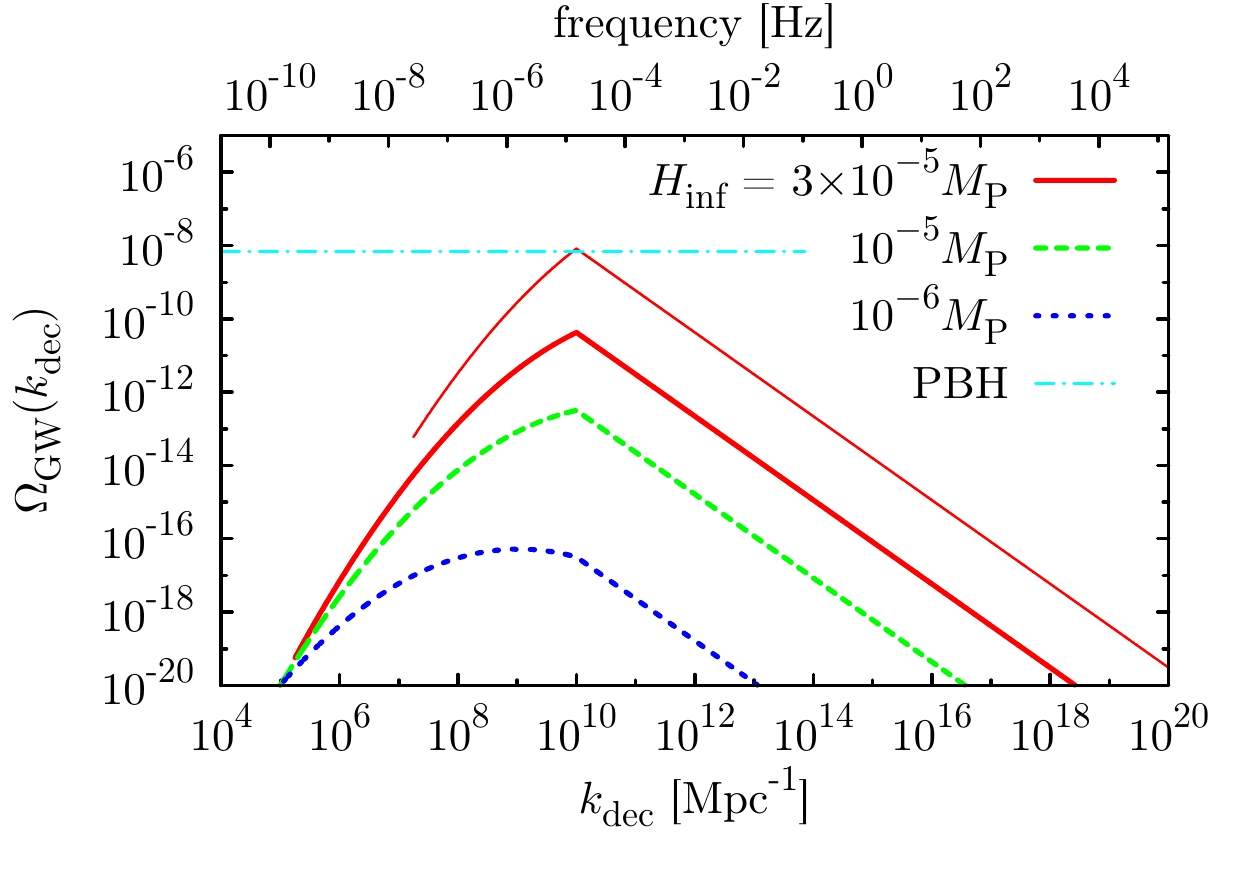}
\label{k_dec_a_1}
}
\subfigure[]{
\includegraphics [width = 7.5cm, clip]{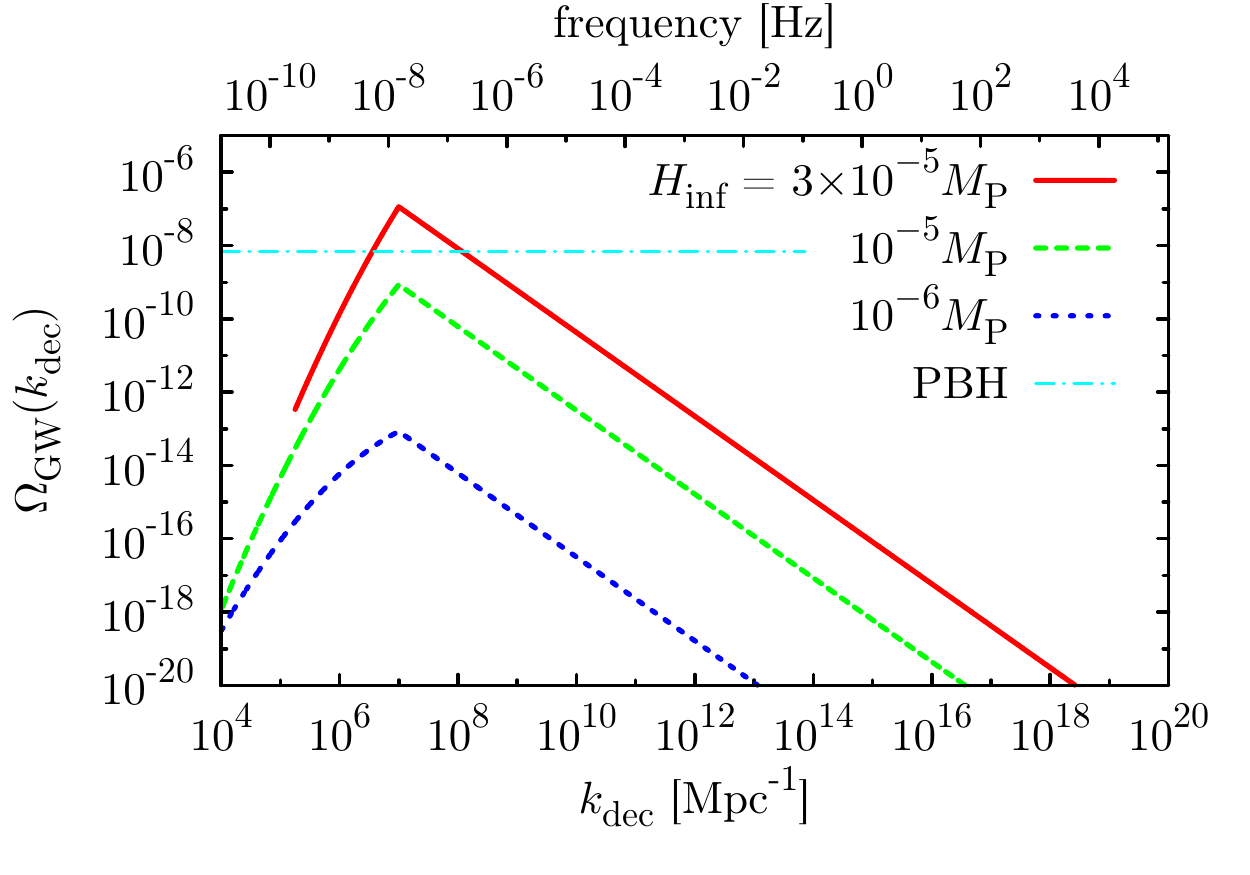}
\label{k_dec_a_2}
}
\caption{
	The peak values of $\Omega_{\rm GW}$ in terms of the peak wave number $k_{\rm dec}$ in axion-like curvaton model are shown.
	We have taken $k_f = 10^{10}~(10^{7})~{\rm Mpc^{-1}}$ for left (right) panel, $r_D = 1$, 
	$H_{\rm inf} = 3 \times 10^{-5}~M_P$ (solid red), $10^{-5}~M_P$ (dashed green) and $10^{-6}~M_P$ (dotted blue), 
	$\kappa = 1$ (thick lines) and $\kappa= 10^{-4}$ (thin lines).
	The dash dotted cyan line shows the upper bound from the PBH overproduction.
}
\label{k_dec_a}
\end{figure}

Next we discuss the axion-like curvaton model.
Fig.~\ref{n_sigma_a} and Fig.~\ref{k_dec_a} show the peak of $\Omega_{\rm GW}$ in terms of $n_\sigma$ and $k_{\rm dec}$ respectively.
We have used the approximated formula (\ref{approx_Omega}) and taken $k_f = 10^{10}~(10^{7})~{\rm Mpc}^{-1}$ in the left (right) panel 
and the same value of $H_{\rm inf}$ for each line as in the quadratic case.
The dash dotted cyan lines corresponds to the upper bound from the PBH overproduction.
Note that the large number of PBHs can be formed for $k_f = 10^{7}~{\rm Mpc}^{-1}$, which may explain the observed dark matter \cite{Kawasaki:2012wr}.
Since we have fixed $r_D$, $H_{\rm inf}$ and $k_f$ as well as $k_c$ and $\mathcal{P}_{\zeta,{\rm curv}}(k_c)$ on each line in Fig.~\ref{n_sigma_a} and Fig.~\ref{k_dec_a}, 
we can see that $k_{\rm dec}$ decreases as $n_\sigma$ increases from Eqs.~(\ref{r_D_1}), (\ref{Gamma_sigma}), (\ref{P_zeta_k_f}) and (\ref{eq:k_dec}).
Thus, the maximum of each line corresponds to $k_{\rm dec} = k_f$ 
because $\mathcal{P}_{\zeta,{\rm curv}}(k_{\rm dec}) = \mathcal{P}_{\zeta,{\rm curv}}(k_c) (k_f/k_c)^{n_\sigma-1}$ for $k_{\rm dec} > k_f$ 
and $\mathcal{P}_{\zeta,{\rm curv}}(k_{\rm dec}) = \mathcal{P}_{\zeta,{\rm curv}}(k_c) (k_{\rm dec}/k_c)^{n_\sigma -1}$ for $k_{\rm dec} < k_f$.
In addition, Fig.~\ref{k_dec_a} shows the maximal amount of gravitational waves increases for small $\kappa$.
It is because the decay rate of the curvaton is inversely proportional to $\kappa$ 
as we can see by fixing $r_D$, $H_{\rm inf}$, $k_f$ and $n_\sigma$.

\begin{figure}[tp]
\centering
\subfigure[]{
\includegraphics [width = 7.5cm, clip]{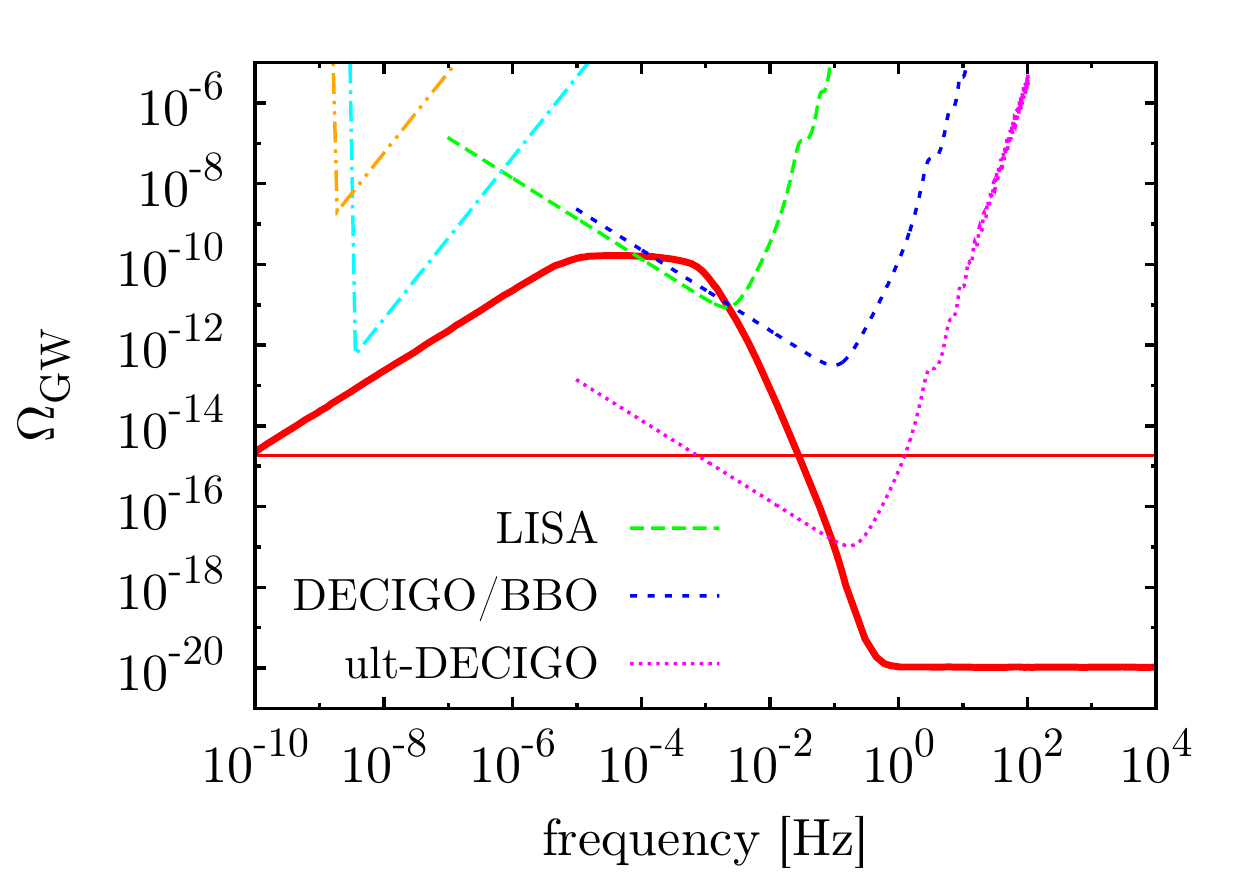}
\label{Omega_GW_axi_1}
}
\subfigure[]{
\includegraphics [width = 7.5cm, clip]{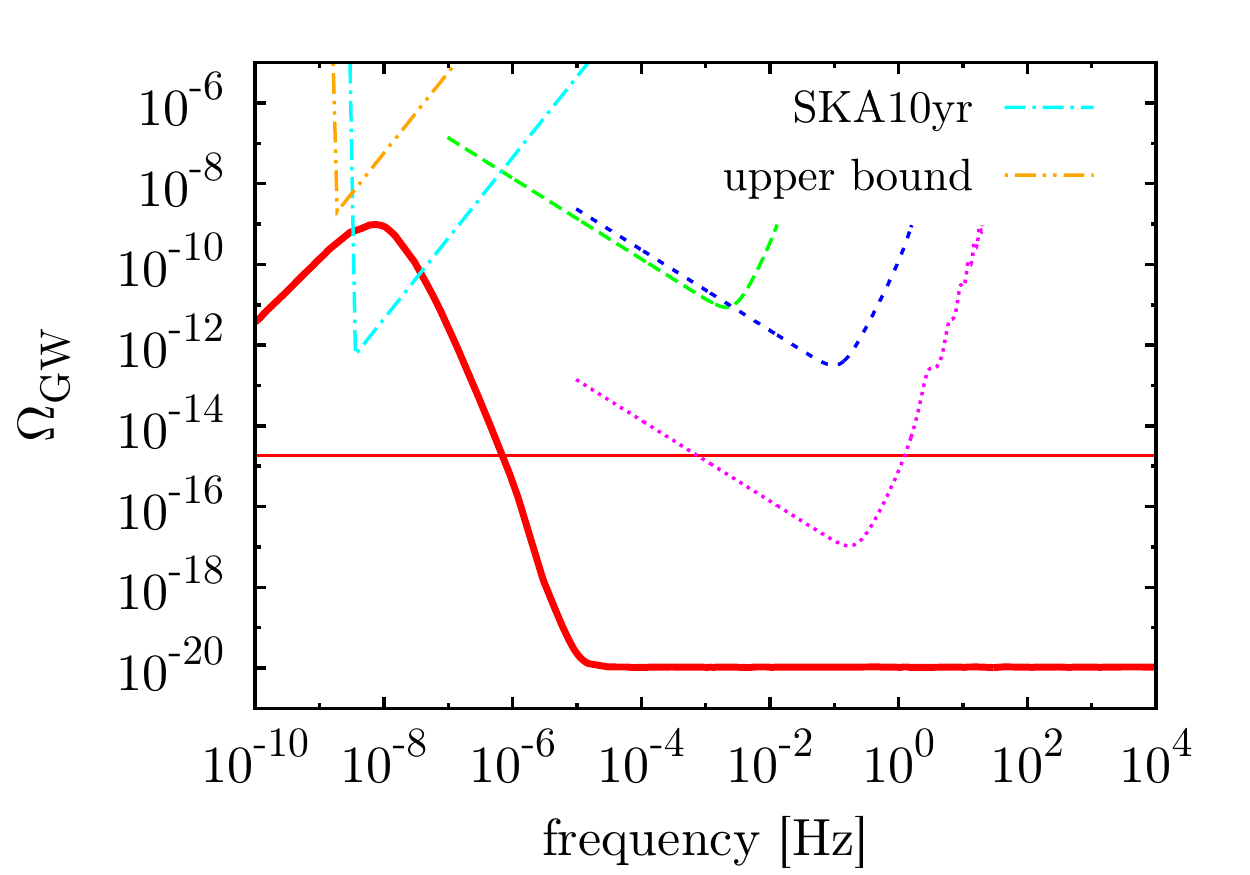}
\label{Omega_GW_axi_2}
}
\caption{
	The spectrum of $\Omega_{\rm GW}$ in an axion-like curvaton model is shown as the thick solid red lines.
	We have taken $H_{\rm inf} = 3\times10^{-5}M_P$, $n_\sigma = 1.5~(1.8)$, 
	$\kappa = 10^{-4}~(1)$, $r_D = 1$ and $k_f = 10^{10}~(10^{7})~{\rm Mpc^{-1}}$ in the left (right) panel. 
	Thin lines shows the contributions from the primordial gravitational waves.
	We show the same sensitivity curves as in Fig.~\ref{Omega_q}.
}
\label{Omega_a}
\end{figure}

Fig.~\ref{Omega_a} shows the resultant spectrum of gravitational waves.
In this figure, we have taken $H_{\rm inf} = 3 \times 10^{-5}M_P$, $n_\sigma = 1.5~(1.8)$, $\kappa = 10^{-4}~(1)$, $r_D = 1$ 
and $k_f = 10^{10}~{\rm Mpc}^{-1}$ in the left (right) panel.
Model parameters are derived to be $f \simeq 2 \times 10^{14}~{\rm GeV}$ ($4 \times 10^{13}~{\rm GeV}$) 
and $m_\sigma \simeq 3 \times 10^{10}~{\rm GeV}$ ($9 \times 10^{3}~{\rm GeV}$) in the left (right) panel.
From this result, we can find that the induced gravitational waves in an axion-like curvaton model can be detectable 
by the future observations such as LISA, DECIGO/BBO or SKA.
In addition, the spectrum has a plateau near the peak, which can be a characteristic signature of the axion-like curvaton model.
Note also that an enhancement due to the curvaton domination can occur in this model if $r_D > 1$ is chosen and the spectrum is raised similar to Fig.~\ref{Omega_q_2}.

\subsection{Contribution from $\mathcal{S}_{ij}^{\rm kin}$}

Next, we focus on the contribution from  $\mathcal{S}_{ij}^{\rm kin}$ given by Eq. (\ref{eq:skin}).
According to \cite{Bartolo:2007vp}, the maximal amount of gravitational waves is estimated as
\be
	\Omega_{\rm GW} \sim 10^{-19} \frac{\Gamma_\sigma}{r_D^2 m_\sigma} \bigg( \frac{\mathcal{P}_\zeta}{2 \times 10^{-9}} \bigg)^2.
	\label{Omega_GW_dsigma}
\ee
Although in \cite{Bartolo:2007vp} the authors have considered a curvaton scenario where
the adiabatic curvature perturbation is generated only from the fluctuation of the curvaton
and assumed the almost scale-invariant power spectrum of the curvature perturbations,
by performing the numerical calculation
we confirm that this estimation is also valid in the blue-tilted curvaton scenario.
As mentioned in Ref. \cite{Bartolo:2007vp},
the above expression indicates that it could be possible to generate the detectable amplitude of the gravitational
waves in the curvaton scenario by tuning model parameters (e.g., $r_D$ and ${\Gamma_\sigma / m_\sigma}$) properly.
However, we show that this contribution is negligible once we take into account the upper bound on the Hubble parameter during inflation.
Defining $X = {\rm min}(m_\sigma,\Gamma_\phi)$, from (\ref{r_D_1}) and (\ref{r_D_2}), we get 
\be
	\frac{\Gamma_\sigma}{r_D^2 m_\sigma} = \frac{1}{36 r_D^4} \bigg( \frac{\sigma_{\rm osc}}{M_P} \bigg)^4 \bigg( \frac{X}{m_\sigma} \bigg)
	\label{enhance_factor}
\ee
which indicates that we need $\sigma_{\rm osc} \gtrsim r M_P$ to enhance the amount of gravitational waves.
Substituting (\ref{power_curv}) and (\ref{enhance_factor}) into (\ref{Omega_GW_dsigma}), we obtain
\be
	\Omega_{\rm GW} \sim 10^{-25} \bigg( \frac{4}{4+3r_D} \bigg)^4 \bigg( \frac{\sigma_{\rm osc}}{\sigma(k)} \bigg)^4 \bigg( \frac{X}{m_\sigma} \bigg) 
	\bigg( \frac{H_{\rm inf}}{10^{14}~{\rm GeV}} \bigg)^4,
\ee
so the amount of gravitational waves cannot be enhanced as long as $\sigma_{\rm osc} < \sigma(k)$ 
even if the curvaton has blue-tilted spectrum.
We check this argument numerically for the quadratic curvaton model and the axion-like curvaton model which we introduced above.

Note that here we consider the contribution of the adiabatic scalar curvature perturbations and the anisotropic stress
due to the kinetic term of the curvaton, separately.
Actually, they have the same origin of the fluctuation of the curvaton and hence
when we consider these two contributions simultaneously
there is a contribution of the cross-term of $\Phi$ and $\delta \sigma$ in the power spectrum of the gravitational
waves. 
However, we confirm that the contribution of the anisotropic stress is subdominant compared with that
of the adiabatic scalar curvature perturbations and can expect that the contribution of such cross-term
is also subdominant.
Furthermore, in this paper we evaluate $\delta \sigma$ on flat slicing but $\Phi$ in the conformal Newtonian gauge.
Thus, there exists the effect of the gauge transformation of $\delta \sigma$ from
flat slicing to the conformal Newtonian gauge when the both contributions are simultaneously included. 
Such an issue to treat the effect of the gauge transformation on second order tensor perturbations explicitly
is left as a future work.

\section{Conclusion}\label{conc}

In this paper, we have considered the curvaton model with two realistic setups, a curvaton model with quadratic potential and an axion-like curvaton model, 
having a blue-tilted power spectrum of curvature perturbation.
In such models, the large curvature perturbation is realized on small scales inaccessible by the CMB temperature anisotropy probe.
Large curvature perturbation induces the large amount of gravitational waves at second order,
so we have calculated the amount of induced gravitational waves in those two curvaton models and 
shown that those models can be testable by the future space-based gravitational wave detectors planned 
such as LISA, DECIGO and BBO or the pulsar timing observation by SKA.  
We have also presented the resultant power spectrum of gravitational waves having a characteristic shape, 
which has a peak at the wave number corresponding to the scale reentering the horizon at the curvaton decay or at the curvaton domination.
Furthermore, in our model, the primordial black holes can be simultaneously formed and account for the cold dark matter.

\section*{Acknowledgment}

We thank Masahide Yamaguchi and Jun'ichi Yokoyama for helpful comments and discussions.
This work is supported by Grant-in-Aid for Scientific research from
the Ministry of Education, Science, Sports, and Culture (MEXT), Japan,
No.\ 22540267 (M.K.), No.\ 21111006 (M.K.) and also 
by World Premier International Research Center
Initiative (WPI Initiative), MEXT, Japan. 
N.K. and S.Y. are supported by the Japan Society for the Promotion of Science (JSPS).

\appendix
\section{Gravitational potential before the curvaton decay} \label{app:grav_pot}

Here we show the analytical expressions for $\Delta_{\rm inf}$ and $\Delta_\sigma$.
Adopting the e-folding number $N$ as a time variable, the differential equation (\ref{evol_Phi}) is rewritten as
\be
	\frac{d\Phi}{dN} + \frac{6+5r}{2+2r} \Phi + \frac{4+3r}{2+2r}\zeta_{\rm inf} + \frac{r}{2+2r} S_\sigma = 0.
\ee
Denoting $r = r_D e^N$ and choosing the initial condition $\Phi(N \to -\infty) = -2\zeta_{\rm inf}/3$ 
which is consistent with the standard (non-curvaton) scenario, 
we obtain the analytic solution for the above first order differential equation:
\be
	\begin{split}
		\Phi &= \bigg( \frac{1}{r^3} -\frac{16\sqrt{1+r}}{15r^3} +\frac{1+r}{15r^3} + \frac{7(1+r)}{15r^2}-\frac{3(1+r)}{5r} \bigg) \zeta_{\rm inf} \\[1mm]
		&~~~ + \bigg( -\frac{1}{r^3} + \frac{16\sqrt{1+r}}{5r^3}-\frac{11(1+r)}{5r^3}+\frac{3(1+r)}{5r^2}-\frac{1+r}{5r} \bigg) S_\sigma, 
	\end{split}
\ee
from which $\Delta_{\rm inf}$ and $\Delta_\sigma$ is expressed explicitly as
\be
	\Delta_{\rm inf} = -1-\frac{6+5r}{4+3r}\bigg( \frac{1}{r^3} -\frac{16\sqrt{1+r}}{15r^3} +\frac{1+r}{15r^3} +\frac{7(1+r)}{15r^2}-\frac{3(1+r)}{5r} \bigg),
\ee
\be
	\Delta_\sigma = 1+\frac{6+5r}{r} \bigg( -\frac{1}{r^3} + \frac{16\sqrt{1+r}}{5r^3}-\frac{11(1+r)}{5r^3}+\frac{3(1+r)}{5r^2}-\frac{1+r}{5r} \bigg).
\ee

\section{Calculation for the power spectrum of gravitational waves} \label{app:power}

In this appendix, we show an explicit calculation for the power spectrum of gravitational waves.
Substituting the superhorizon evolution for $\Phi$, given by (\ref{deviation}), 
and the transfer function, given by (\ref{transfer_Phi}), (\ref{transfer_sigma_1}) and (\ref{transfer_sigma_2}), 
into (\ref{source_term_Phi}) and (\ref{source_term_kin}), we obtain the following expression for the contribution from curvaton:
\be
	\mathcal{S}^i ({\bf k},\eta) = \int \frac{d^3k}{(2\pi)^{3/2}}e({\bf k}, \tilde{\bf k}) 
	f_i ({\bf k}, \tilde{\bf k},\eta) S_{\sigma}(\tilde{\bf k}) S_{\sigma}({\bf k}-\tilde{\bf k})~~~\text{with}~~~ i = \Phi,~{\rm kin},
\ee
where we define
\be
	e({\bf k}, \tilde{\bf k}) = e^{lm}({\bf k}) \tilde{k}_l \tilde{k}_m = \sqrt{2} \tilde{k}^2 \sin^2\theta
\ee
and 
\begin{align}
	\begin{split}
		f_\Phi ({\bf k},\tilde{\bf k},\eta) =&~ 4\bigg( \frac{r(1-\Delta_\sigma)}{6+5r} \bigg)^2 
		\bigg[ \bigg( 3+\frac{2\Delta_\sigma}{(1-\Delta_\sigma)^2} \bigg( \frac{6+5r}{2+2r} \bigg) \\[1mm]
		&~~~~~~+\bigg( \frac{6+5r}{2+2r} \bigg)^2 \bigg( \frac{\Delta_\sigma}{1-\Delta_\sigma} \bigg)^2  \bigg) 
		T_\Phi(\tilde{k},\eta) T_\Phi(|{\bf k}-\tilde{\bf k}|,\eta) \\[2mm]
		&~~~ +\frac{2}{\mathcal{H}} \bigg( 1+ \frac{\Delta_\sigma}{1-\Delta_\sigma} \bigg(\frac{6+5r}{2+2r} \bigg) \bigg) 
		T'_\Phi(\tilde{k},\eta) T_\Phi(|{\bf k}-\tilde{\bf k}|,\eta) \\[1mm]
		&~~~ + \frac{1}{\mathcal{H}^2} T'_\Phi(\tilde{k},\eta) T'_\Phi(|{\bf k}-\tilde{\bf k}|,\eta) \bigg],
	\end{split} \\[1mm]
	f_{\rm kin}({\bf k},\tilde{\bf k},\eta) = & ~ - \bigg( \frac{\sigma_{\rm osc}}{M_P} \bigg)^2 
	T_{\delta\sigma}(\tilde{k},\eta) T_{\delta\sigma}(|{\bf k}-\tilde{\bf k}|,\eta) \theta(1-k_{\rm dec}\eta).
\end{align}
Then, substituting these source terms into (\ref{P_h}), 
and assuming that the primordial isocurvature perturbation, $S_\sigma({\bf k})$, obeys the Gaussian statisctics,  
we obtain 
\be
	\begin{split}
		\langle h_{\bf k}(\eta) h_{\bf p}(\eta) \rangle 
		&= \frac{2\pi^2}{k^3} \delta^3({\bf k} + {\bf p}) \int^{\infty}_0 dy \int^{1+y}_{|1-y|} dx \\[1mm]
		&~~~ \times \frac{2y^2}{x^2} \bigg( 1-\frac{(1+y^2-x^2)^2}{4y^2} \bigg)^2 
		\mathcal{P}_{S,{\rm curv}}(kx) \mathcal{P}_{S,{\rm curv}}(ky) \\[1mm]
		&~~~ \times \bigg[ \frac{k^2}{a(\eta)} \int^{\eta}_{\eta_0} d\tilde\eta a(\tilde\eta)
		g_{\bf k}(\eta;\tilde\eta) f_i (kx,ky,\tilde{\eta}) \bigg]^2,
	\end{split}
\ee
where we define $x = |{\bf k}-\tilde{\bf k}|/k$ and $y = \tilde{k}/k$.
This leads the power spectrum of gravitational waves through (\ref{P_h}).
We have calculated this integral numerically to yield the resultant spectrum shown in Fig.~\ref{Omega_q} and Fig.~\ref{Omega_a}.

\end{document}